\documentclass[article,english,12pt]{article}

\usepackage{amscd,amsfonts,amsmath,amssymb,amsthm,bbm,bm,latexsym,mathrsfs}
\usepackage{epsfig,graphics,graphicx,natbib,subfigure,longtable}
\usepackage[english]{babel}
\usepackage[usenames]{color}
\usepackage{bookmark}
\usepackage{sgame}
\usepackage{gensymb}
\usepackage{epstopdf}
\usepackage{float}
\usepackage{comment}
\usepackage{geometry}
\usepackage{appendix}
\usepackage{amsmath}
\usepackage{lscape}
\usepackage{bbm}
\usepackage[super]{nth}

\usepackage{tikz} 
\usetikzlibrary{matrix}
\usetikzlibrary{bayesnet}
\usetikzlibrary{arrows,shapes,positioning,mindmap,trees,automata}

\makeatother

\usepackage{amsmath}

\theoremstyle{remark}

\theoremstyle{plain}

\begin{document}

\title{\vspace{-50pt} Social Media Integration of Flood Data: A Vine Copula-Based Approach}%\thanks{Authors research is supported by funding from ...}}

\author{
Lauren Ansell \hspace{15pt} and \hspace{15pt}
Luciana Dalla Valle \hspace{15pt} %\vspace{5pt} \\
        \\
        \vspace{-10pt}
        \\
        {\centering {\small
        University of Plymouth}} 
             } 
\date{\today}
\maketitle

%%%%%%%%%%%%%%%%%%%%%%%%%%%%%%%%%%%
%%         	Abstract			%%%
%%%%%%%%%%%%%%%%%%%%%%%%%%%%%%%%%%%
\begin{abstract}

Floods are the most common and among the most severe natural disasters in many countries around the world. As global warming continues to exacerbate sea level rise and extreme weather, governmental authorities and environmental agencies are facing the pressing need of timely and accurate evaluations and predictions of flood risks. Current flood forecasts are generally based on historical measurements of environmental variables at monitoring stations. In recent years, in addition to traditional data sources, large amounts of information related to floods have been made available via social media. Members of the public are constantly and promptly posting information and updates on local environmental phenomena on social media platforms. Despite the growing interest of scholars towards the usage of online data during natural disasters, the majority of studies focus exclusively on social media as a stand-alone data source, while its joint use with other type of information is still unexplored. In this paper we propose to fill this gap by integrating traditional historical information on floods with data extracted by Twitter and Google Trends. Our methodology is based on vine copulas, that allow us to capture the dependence structure among the marginals, which are modelled via appropriate time series methods, in a very flexible way. We apply our methodology to data related to three different coastal locations on the South coast of the United Kingdom (UK). The results show that our approach, based on the integration of social media data, outperforms traditional methods in terms of evaluation and prediction of flood events. 

\vspace{1cm}


%\textbf{JEL Classifications: } {\bf ??????}

 \end{abstract}

%%%%%%%%%%%%%%%%%%%%%%%%%%%%%%%%%%%
%%         		Intro			%%%
%%%%%%%%%%%%%%%%%%%%%%%%%%%%%%%%%%%

\section{Introduction}\label{intro}

In recent years, climate change has caused an exacerbation of the frequency and severity of natural hazard phenomena, such as floods, storms, wildfires and other extreme weather events \citep{field2012managing, muller2015crowdsourcing}.
Around the world, a substantial part of the population is exposed to flood risk, with more than 2.3 billion people residing in locations experiencing inundations during flood events \citep{UN2015}.
In the United Kingdom, intense storms occurred during recent years, bringing severe flooding and causing considerable damage to people, infrastructure and the economy, totalling millions of pounds \citep{smith2017assessing}.
This caused a growing need for timely and accurate information about the severity of flooding, which is essential for forecasting and nowcasting these phenomena and for effectively managing response operations and appropriately allocate resources \citep{rosser2017rapid}. 

Generally, in order to estimate and predict inundations, statistical and machine learning models are employed, typically using information gathered from meteorological and climatological instrumentation at monitoring stations.  
For example, \cite{wang2003internet} use a combination of meteorological, geographical and urban data to produce flooding tables and maps published via Internet for public consultation.
\cite{keef2013estimating} used data from a set of UK river flow gauges to estimate the probability of widespread floods based on the conditional exceedance model of \cite{heffernan2004conditional}.
\cite{grego2015standard} collected historic flood frequency data and modelled them via finite mixture models of stationary distributions using censored data methods.
\cite{balogun2020improved} utilized geographic information system and remote sensing data from Malaysia to generate flood susceptibility maps, applying Fuzzy-Analytic Network Process flood models. Model validation results showed that 59.42\% and 36.23\% of past flood events fall within the very high and high susceptible locations of the susceptibility map respectively. 
\cite{moishin2020development} investigated fluvial flood risk in Fiji developing a flood index based on current and antecedent day’s precipitation.
\cite{talukdar2020flood} gathered historical flood data related to the Teesta River basin in Bangladesh and employed ensemble machine learning algorithms to predict flooding sites and flood susceptible zones. Results showed that an area of more than 800 km$^2$ was predicted as a very high flood susceptibility zone by all algorithms. 

%\cite{wang2003internet} use a combination of meteorological, geographical and urban data to produce flooding tables and map published via Internet for public consultation.
%It first delineates flooding areas in response to different rainfall amounts and land use scenarios, then overlays the flooding areas with other data layers to identify properties and streets potentially to be affected by the user-specified flooding event. Finally, results from the simulation and overlay analysis are published via Internet as tables and maps. 

However, information collected at monitoring stations may suffer from data sparsity, time delays and high costs \citep{muller2015crowdsourcing}. In particular, remotely sensed data may take several hours to become available \citep{mason2012near} and their temporal resolution is often limited \citep{schumann2009progress}.

On the other hand, an increasing availability of consumer devices, such as smartphones and tablets, is leading to the dissemination and communication of flood events directly by individuals, with information shared in real-time using social media. User-generated content shared online often includes reports on meteorological conditions especially in case of extreme or unusual weather \citep{alam2018crisismmd}.
Recent studies have focused specifically on social media sources, such as Twitter, Facebook and Flickr, to collect real-time information on floods and environmental events and their impacts across the globe.
For example, \cite{herfort2014exploring} and \cite{de2015geographic} identified spatial patterns in the occurrence of flood-related tweets associated with proximity and severity of the River Elbe flood in Germany in June 2013.
\cite{saravanou2015twitter} performed a case study on the floods that occurred in the UK during January 2014, investigating how these were reflected on Twitter. The authors evaluated their findings against ground truth data, obtained from external independent sources, and were able to identify flood-stricken areas. 
Twitter data generated during flooding crisis was also used by \cite{spielhofer2016data} to evaluate techniques to be adopted in real-time to provide actionable intelligence to emergency services. 
Different methods to create flood maps from Twitter micro-blogging were presented by \cite{brouwer2017probabilistic}, \cite{smith2017assessing} and \cite{arthur2018social}, who applied their approaches to different locations, such as the city of York (UK), Newcastle upon Tyne (UK) and the whole England region, respectively.
%\cite{brouwer2017probabilistic} presented a method to create flood maps from Twitter micro-blogging that mentions locations of flooding, applying their approach to the December 2015 flood in the city of York (UK).
%In the paper by \cite{smith2017assessing}, the authors illustrated a real-time modelling framework to identify areas in Newcastle upon Tyne (UK) likely to have ﬂooded using data obtained only through social media.
%\cite{arthur2018social} investigated the use of social sensing for observing natural hazards and produce historical and real-time maps of floods using Twitter data.
The 2015 South Carolina flood disaster was analysed by \cite{li2018novel} to map the flood in real time by leveraging Twitter data in geospatial processes. Results show that the authors’ approach could provide a consistent and comparable estimation of the flood situation in near real time. 
\cite{spruce2021social} analysed rainfall events occurred across the globe in 2017, comparing outputs from social sensing against a manually curated database created by the Met Office. The authors showed that social sensing successfully identified most high-impact rainfall events present in the manually curated database, with an overall accuracy of 95\%.

However, the majority of contributions in the literature analysing online generated data focus exclusively on social media sources, overlooking any relation or synergy with other sources of information.
One of the few exceptions is the paper by \cite{rosser2017rapid}, who estimated the flood inundation extent in Oxford (UK) in 2014 based on the fusion of remote sensing, social media and topographic data sources, using a simple Weights-of-evidence analysis.

In this paper we propose to leverage the association between social media and environmental information via sophisticated statistical modelling based on vine copulas, to enhance the assessment and prediction of flood phenomena compared to traditional approaches.

Copulas are multivariate statistical tools, which allow us to model separately the marginal models and their dependence structure \citep{huang2017modelling}.
Copulas were used in flood risk analysis, for example,
by \cite{jane2016copula} to predict the wave height at a given location by exploiting the spatial dependence of the wave height at nearby locations.
The use of copulas in flood risk management was also explored by \cite{jane2018exploring}, who used a copula to capture dependencies in a 3-dimensional loading parameter space, estimating the overall failure probability.
Copulas have also been applied in a flood risk context to model the dependence between multiple co-occurring drivers by \cite{ward2018dependence}, among others. \cite{couasnon2018copula} use Gaussian pair-copulas in a Bayesian Network to derive boundary conditions that account for riverine and coastal interactions for a catchment in southeast Texas. \cite{feng2020nonstationary} employed time-varying copulas with nonstationary marginal distributions to estimate the dependence structure of inundation magnitudes in flood coincidence risk assessment.

Vine copulas are based on bivariate copulas as building blocks and provide a great deal of flexibility, compared to standard copulas and other traditional multivariate approaches, in modelling complex dependence structures between the variables.
Vine copulas were adopted, for example, by \cite{latif2020parametric} to model trivariate flood characteristics for the Kelantan River basin in Malaysia. 
\cite{tosunoglu2020multivariate} applied vine copulas in hydrology for multivariate modelling of peak, volume and duration of floods in the Euphrates River Basin, Turkey.
Vine copulas were applied to model compound events by \cite{bevacqua2017multivariate} and by \cite{santos2021assessing}. The former authors adopted this approach to quantify the risk in present-day and future climate, and to measure uncertainty estimates around such risk. The latter authors used vines to assess compound flooding from storm surge and multiple riverine discharges in Sabine Lake, Texas.

However, to the best of our knowledge, there are currently no studies exploring the use of vine copulas to integrate social media data with other types of information. This paper proposes a novel approach, based on vine copulas, that combines data gathered from Twitter and Google Trends with remotely sensed information. The proposed methodology involves the use of subjective information, more specifically the feelings of people expressed through social media and quantified by sentiment scores, not merely as stand-alone data sources, used in isolation to predict inundations, but combined with information on the occurrence and magnitude of flood events. The vine copula approach allows us to exploit the associations between all the considered data sources, environmental as well as on-line, which all contribute to calculate flood forecasts.

The methodology articulates in the following steps, that will be illustrated in detail in the following sections:
\begin{enumerate}
    \item fit each variable (environmental as well as on-line information) with a suitable time series model, to remove the temporal effects from the data;
    \item construct a vine copula model, which accounts for the dependencies between all variables and exploits the associations between environmental and social media information;
    \item calculate predictions of the flood variables based on the vine copula model.
\end{enumerate}
The application of our methodology to three different coastal locations in the South of the UK shows that our approach performs better than traditional approaches, which do not take into account associations between environmental and on-line information, to estimate and predict the occurrence and the magnitude of flood events.

The remainder of the paper is organised as follows. Section \ref{sec:data} describes the environmental and social media data used in the analysis; Section \ref{sec:method} illustrates the vine copula methodology; Section \ref{sec:analysis} reports the results of the analysis; finally, concluding remarks are presented in Section \ref{sec:Conclusions}.

%%%%%%%%%%%%%%%%%%%%%%%%%%%%%%%%%%%
%%     Datasets 			%%%
%%%%%%%%%%%%%%%%%%%%%%%%%%%%%%%%%%%

\section{Study Area and Data Collection}\label{sec:data}

The UK coastline has been subject to terrible floods throughout history. Over the last few years, storms and floods relentlessly hit the UK coast, triggering intense media coverage and public attention. 
Table \ref{tab:storms} lists the major winter storm events affecting the UK between 2012 and 2018.
\begin{table}[htbp]
  \centering
  \caption{Major winter storm events in the UK between 2012 and 2018. Note that the storm naming system was introduced in 2015.}
  \begin{scriptsize}
    \begin{tabular}{|c|c|c|c|c|c|c|c|}
    \hline
    \multicolumn{1}{|c|}{\textbf{Winter}} & \textbf{Winter} & \multicolumn{2}{c|}{\textbf{Winter}} & \multicolumn{2}{c|}{\textbf{Winter}} & \multicolumn{2}{c|}{\textbf{Winter}} \\
    \multicolumn{1}{|c|}{\textbf{2012/13}} & \textbf{2013/14} & \multicolumn{2}{c|}{\textbf{2015/16}} & \multicolumn{2}{c|}{\textbf{2016/17}} & \multicolumn{2}{c|}{\textbf{2017/18}} \\
    \hline
    \multicolumn{1}{|c|}{\textbf{Date}} & \multicolumn{1}{c|}{\textbf{Date}} & \multicolumn{1}{c|}{\textbf{Storm}} & \multicolumn{1}{c|}{\textbf{Date}} & \multicolumn{1}{c|}{\textbf{Storm}} & \multicolumn{1}{c|}{\textbf{Date}} & \multicolumn{1}{c|}{\textbf{Storm}} & \multicolumn{1}{c|}{\textbf{Date}} \\
     &  & \multicolumn{1}{c|}{\textbf{Name}} &  & \multicolumn{1}{c|}{\textbf{Name}} &  & \multicolumn{1}{c|}{\textbf{Name}} &  \\
    \hline
    \multicolumn{1}{|c|}{11 Oct } & 28 Oct  & \multicolumn{1}{c|}{Abigail  } & \multicolumn{1}{c|}{ 12-13 Nov } & \multicolumn{1}{l|}{Angus } & \multicolumn{1}{c|}{ 20 Nov } & \multicolumn{1}{c|}{Aileen } & \multicolumn{1}{c|}{ 12-13 Sep} \\
    \multicolumn{1}{|c|}{18 Nov } & 5-6 Dec & \multicolumn{1}{c|}{ Barney } & \multicolumn{1}{c|}{ 17-18 Nov } & \multicolumn{1}{c|}{ Barbara } & \multicolumn{1}{c|}{ 23-24 Dec } & \multicolumn{1}{c|}{ Brian } & \multicolumn{1}{c|}{ 21 Oct } \\
    \multicolumn{1}{|c|}{14 Dec } & 18-19 Dec & \multicolumn{1}{c|}{ Clodagh } & \multicolumn{1}{c|}{ 29 Nov } & \multicolumn{1}{c|}{ Conor } & \multicolumn{1}{c|}{ 25- 26 Dec } & \multicolumn{1}{c|}{ Caroline } & \multicolumn{1}{c|}{ 7 Dec } \\
    \multicolumn{1}{|c|}{19 Dec } & 23-24 Dec  & \multicolumn{1}{c|}{ Desmond } & \multicolumn{1}{c|}{ 5-6 Dec  } & \multicolumn{1}{c|}{ Doris } & \multicolumn{1}{c|}{ 23 Feb } & \multicolumn{1}{c|}{ Dylan } & \multicolumn{1}{c|}{ 30-31 Dec } \\
    \multicolumn{1}{|c|}{22 Dec } & 26-27 Dec & \multicolumn{1}{c|}{ Eva } & \multicolumn{1}{c|}{ 24 Dec } & \multicolumn{1}{c|}{ Ewan } & \multicolumn{1}{c|}{26 Feb} & \multicolumn{1}{c|}{ Eleanor } & \multicolumn{1}{c|}{ 2-3 Jan } \\
          & 30-31 Dec & \multicolumn{1}{c|}{ Frank } & \multicolumn{1}{c|}{ 29-30 Dec } &       &       & \multicolumn{1}{c|}{ Fionn } & \multicolumn{1}{c|}{ 16 Jan } \\
          & 3 Jan & \multicolumn{1}{c|}{ Gertrude } & \multicolumn{1}{c|}{ 29 Jan } &       &       & \multicolumn{1}{c|}{ Georgina } & \multicolumn{1}{c|}{24 Jan} \\
          & 25-26 Jan  & \multicolumn{1}{c|}{ Henry } & \multicolumn{1}{c|}{ 1-2 Feb } &       &       &       &  \\
          & 31 Jan-1 Feb  & \multicolumn{1}{c|}{ Imogen } & \multicolumn{1}{c|}{ 8 Feb } &       &       &       &  \\
          & 4-5 Feb  & \multicolumn{1}{c|}{ Jake } & \multicolumn{1}{c|}{ 2 Mar } &       &       &       &  \\
          & 8-9 Feb  & \multicolumn{1}{c|}{ Katie } & \multicolumn{1}{c|}{ 27-28 Mar } &       &       &       &  \\
          & 12 Feb  &       &       &       &       &       &  \\
          & 14-15 Feb  &       &       &       &       &       &  \\
    \hline
    \end{tabular}%
    \end{scriptsize}
  \label{tab:storms}%
\end{table}%

In this paper we consider three locations on the South coast of the UK, which were severely affected by storm events in recent years: Portsmouth, Plymouth and Dawlish. The inundation episodes of the last few years had a substantial socioeconomic impact on the local communities of the three locations, which are totalling a population of almost 500,000. 
The three areas were affected by most of the inundation events listed in Table \ref{tab:storms}.
In particular, devastating overnight storms on February 4, 2014, swept the main rail route at Dawlish, leaving tracks dangling in mid-air. The seawall was breached, a temporary line of shipping containers forming a breakwater was constructed, however huge waves damaged it and punched a new hole in the sea wall. Later, a replacement seawall was installed and railway operations re-commenced on April 4, 2014. The waves on the night of the \nth{4} February were relatively modest. The breach was more likely a result of a combination of factors including coincidental arrival of swell waves and the highest locally generated wind waves, large storm surge arriving a few days after a spring tide and the sequence of storm events hitting the South UK coast that winter before the breach lowering beach level \citep{sibley2015coastal}.

In order to estimate and predict flood phenomena in the three coastal areas, we applied the vine copula methodology to data based on historical measurement in conjunction with information gathered online.

For each one of the three locations, we obtained daily hydraulic loading condition data as well as social media information for the period between January 2012 and December 2016, obtaining $1,827$ daily data points for each variable. We therefore constructed a dataset of time series, all of the same length.
More precisely, we downloaded wave height (m) and water level (tidal residual, m) data from the UK Environment Agency flood-monitoring API \footnote{Available at the website \url{https://environment.data.gov.uk/flood-monitoring/doc/reference}}. 
Furthermore, for the aforementioned locations, we gathered Google Trends information on the number of searches for the keywords \textit{flood}, \textit{flooding}, \textit{rain} and \textit{storm}, using the \texttt{gtrendsR} package from the R software \citep{gtrendsR, Rsoftware}.
In addition, we collected Twitter messages containing the same keywords used to perform Google Trends searches for the three areas. After removing tweets sent by automated accounts, which contained factual information about the current weather in the required location, we obtained 9,781 tweets for Portsmouth, 4,995 tweets for Plymouth and 1,769 tweets for Dawlish.
From the Twitter data, we considered the total number of tweets as well as the sentiment scores calculated using two different lexicons: Bing and Afinn \citep{hu2004mining}, which are available in the R \texttt{tidytext} package \citep{tidytext}.
The Bing lexicon splits words into positive or negative. The Bing sentiment score for each tweet is calculated by counting the number of positive words used in each tweet and subtracting from this the number of negative words.
The Afinn lexicon scores words between $\pm 5$. The Afinn sentiment score is calculated by multiplying the score of each word by the number of times it appears in the tweet; these scores are then summed to
derive the overall sentiment score.
In order to take into account of the different population sizes living in the three areas \footnote{We considered a total population of 238,137 for Portsmouth; a total population of 234,982 for Plymouth; a total population of 16,298 for Dawlish. \textit{Source}: 2011 United Nations population figure, available at: \url{https://unstats.un.org/unsd/demographic-social/}}, we scaled the Bing and Afinn sentiment scores by the relevant number of residents.

\begin{figure}[h]
    \centering
    \includegraphics[width=0.9\textwidth]{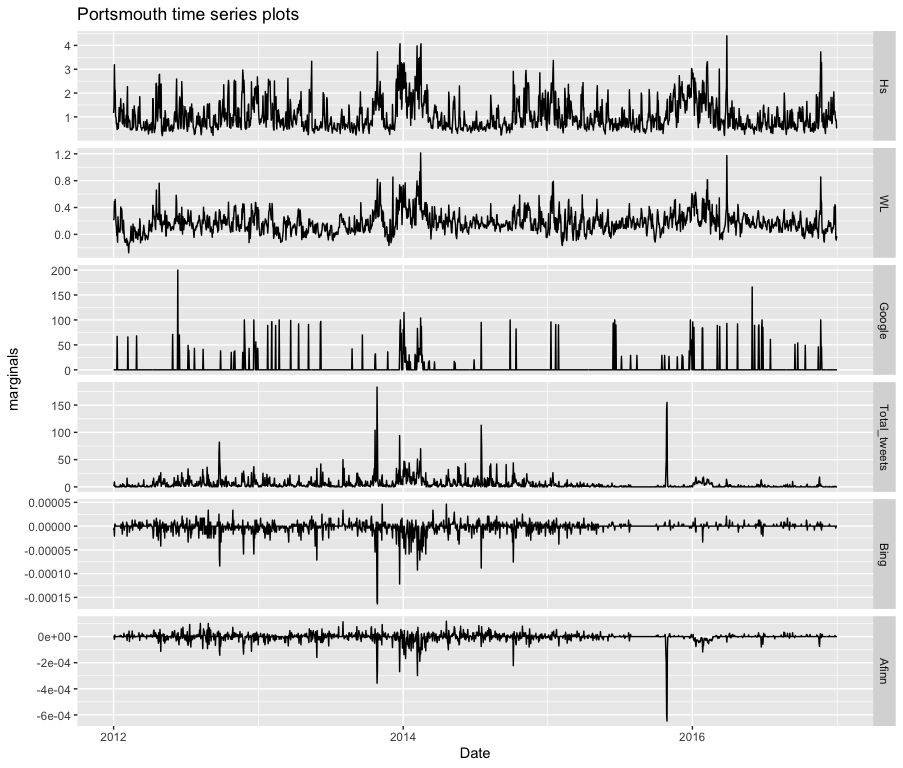} 
\caption{Trace plots of Portsmouth data.}
    \label{fig:Port_trace}
\end{figure}    
    
\begin{figure}[h]
    \centering    
    \includegraphics[width=0.9\textwidth]{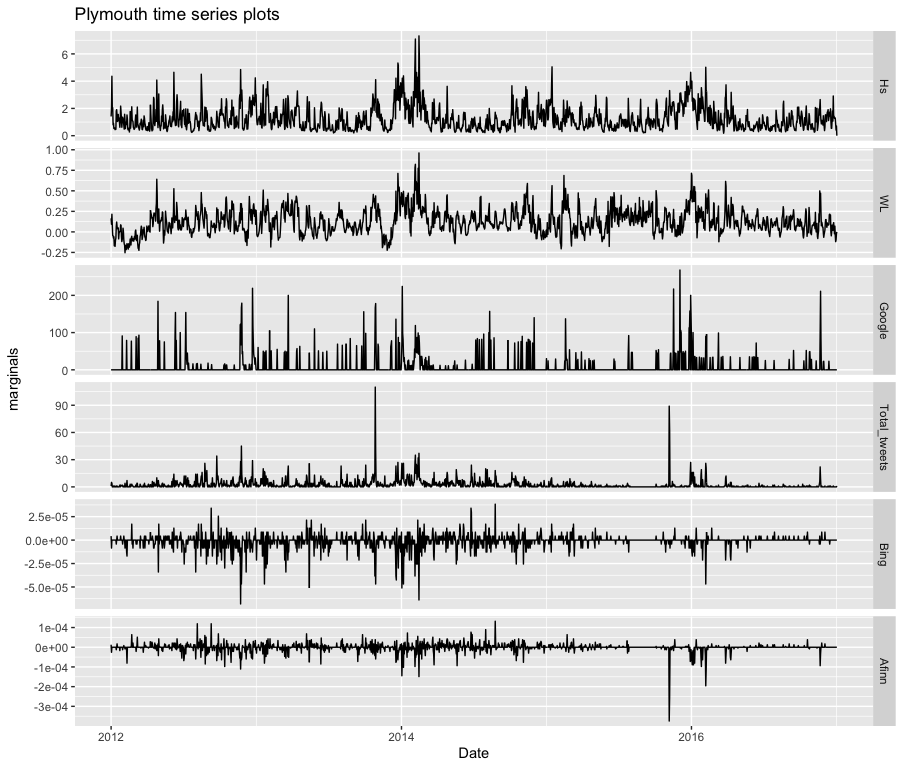} 
    \caption{Trace plots of Plymouth data.}
    \label{fig:Plym_trace}
\end{figure}

\begin{figure}[h]
    \centering  
    \includegraphics[width=0.9\textwidth]{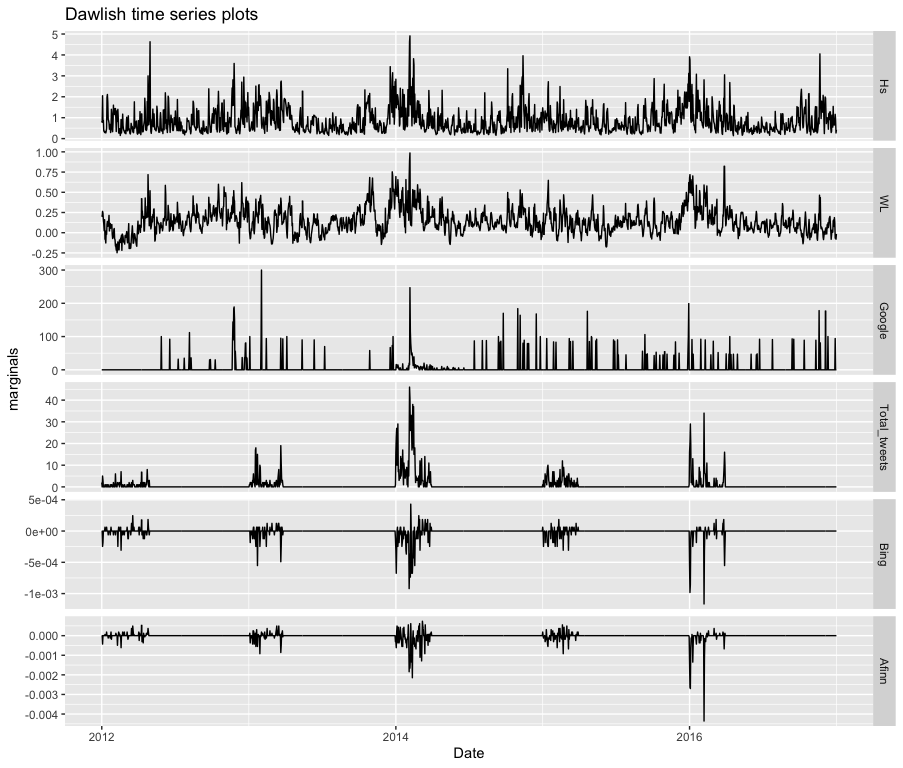}
    \caption{Trace plots of Dawlish data.}
    \label{fig:Dawl_trace}
\end{figure}

Figures \ref{fig:Port_trace}, \ref{fig:Plym_trace} and \ref{fig:Dawl_trace} show the trace plots of the data collected for Portsmouth, Plymouth and Dawlish, respectively.
The plots are produced using a daily scale. 
The panels (from top to bottom) illustrate the wave height (\texttt{Hs}), the water level (\texttt{WL}), the Google Trends searches (\texttt{Google}), the total number of Tweets (\texttt{Total\_tweets}), the Bing sentiment scores (\texttt{Bing}) and the Afinn sentiment scores (\texttt{Afinn}).
We notice spikes in the plots corresponding to most of the storm events listed in Table \ref{tab:storms}. For example, the flood events occurred in February 2014 are reflected in high spikes in the time series plots, especially for Dawlish in Figure \ref{fig:Dawl_trace}. 
From the plots we also notice that the time series exhibit a similar pattern at specific time points. Generally, the higher the values of wave height and water level, the higher the volume of tweets and Google searches, and the lower the sentiment scores for both lexicons. This suggest the presence of association between the social media and remotely-sensed data.

%%%%%%%%%%%%%%%%%%%%%%%%%%%%%%%%%%%
%%     Methodology 			%%%
%%%%%%%%%%%%%%%%%%%%%%%%%%%%%%%%%%%

\section{Methodology}\label{sec:method}

The copula is a function that allows us to bind together a set of marginals, to model
their dependence structure and to obtain the joint multivariate distribution \citep{joe1997multivariate, nelsen2007introduction}.
Sklar’s theorem \citep{sklar1959fonctions} is the most important result in copula theory. It states that, given a vector of random variables $\textbf{X}=(X_1, \ldots, X_d)$, with $d$-dimensional joint cumulative distribution function $F(x_1, \ldots,x_d)$ and marginal cumulative distributions (cdf) $F_j(x_j)$, with $j=1, \ldots, d$, a $d$-dimensional copula $C$ exists, such that
$$
F(x_1, \ldots,x_d) = C(F_1(x_1), \ldots, F_d(x_d); \boldsymbol{\theta}),
$$
where $F_j(x_j) = u_j$, with $u_j \in [0,1]$ are called \textit{u-data}, and $\boldsymbol{\theta}$ denotes the set of parameters of the copula.
The joint density function can be derived as
$$
f(x_1, \ldots,x_d) = c(F_1(x_1), \ldots, F_d(x_d); \boldsymbol{\theta}) \cdot f_1(x_1) \cdot \cdot \cdot f_d(x_d),
$$
where $c$ denotes the $d$-variate copula density.
The copula allows us to determine the joint multivariate distribution and to describe the dependencies among the marginals, that can potentially be all different and can be modelled using distinct distributions.

In this paper, we adopt the 2-steps inference function for margins (IFM) approach \citep{joe1996estimation}, estimating the marginals in the first step, and then the copula, given the marginals, in the second step.

%%%%%%%%%%%%%%%%%%%%%%%%%%%%%%%%%%%
%%     Marginal Models 			%%%
%%%%%%%%%%%%%%%%%%%%%%%%%%%%%%%%%%%

\subsection{Marginal Models}\label{sub:marginals}

Given the different characteristics of the six marginals, we fitted different models for each of the six time series for each location.
Further, we extracted the residuals $\varepsilon_j$, with $j = 1, \ldots, d$, from each marginal model and we applied the relevant distribution functions to get the \textit{u-data} $F_j(\varepsilon_j) = u_j$ to be plugged into the copula.
%According to the characteristics of each of the six time series marginals for each location, we used different models which provided the best fit to the data.
%Then, the \textit{u-data} to be plugged into the copula were obtained first extracting the residuals $\varepsilon_j$, with $j = 1, \ldots, d$, from each marginal model and then applying the relevant distribution function $F_j(\varepsilon_j) = u_j$.

\subsubsection{Wave height (\texttt{Hs})}

The best fitting model for the log-transformed \texttt{Hs} marginal for all three locations was the autoregressive integrated moving average (ARIMA) model (for more information about ARIMA models, see, for example \cite{hyndman2018forecasting}). 
The ARIMA model aims to describe
the autocorrelations in the data by combining autoregressive and moving average models. The model is usually denoted as ARIMA($p$, $d$, $q$), where the values in the brackets indicate the parameters: $p$, $d$, $q$, where $p$ is the order of the autoregressive part, $d$ is the degree of first differencing involved
and $q$ is the order of the moving average part.
The ARIMA model, for $t = 1,\ldots, T$ takes the following form:
\begin{equation}\label{eq:ARIMA}
    y_t = a + \sum_{i=1}^p \phi_i y_{t-i} + \sum_{i=1}^q \theta_i \varepsilon_{t-i} + \varepsilon_t
\end{equation}
where $y_t =(1-B)^d x_t$, $x_t$ are the original data values, $B$ is the backshift operator,
$a$ is a constant, $\phi_i$, with $i = 1, \ldots, p$, are the autoregressive parameters, $\theta_i$, with $i =1,\ldots,q$, are the moving average parameters and $\varepsilon_t \sim N(0,1)$ is the error term.

\subsubsection{Water level (\texttt{WL})}

We fitted the log-transformed \texttt{WL} marginal for the Plymouth location with an ARIMA model, as described in Eq.\eqref{eq:ARIMA}.
However, for Portsmouth and Dawlish, the ARIMA-GARCH model with Student's t innovations appeared to have a better fit.
This model combines the features of the ARIMA model with the generalized autoregressive conditional heteroskedastic (GARCH) model, allowing us to capture time series volatility over time. The GARCH model is typically denoted as GARCH($p$, $q$), with parameters $p$ and $q$, where $p$ is the number of lag residuals errors and $q$ is the number of lag variances.
The ARIMA($p$, $d$, $q$)-GARCH($p$, $q$) model can be expressed as:
$$
    y_t = a + \sum_{i=1}^p \phi_i y_{t-i} + \sum_{i=1}^q \theta_i \varepsilon_{t-i} + \varepsilon_t 
$$
\begin{equation}\label{eq:ARIMA-GARCH}
    \varepsilon_t = \sqrt{\sigma_t} z_t \hspace{1cm} \sigma^2 = \omega + \sum_{i=1}^p \alpha_i \varepsilon_{t-i}^2 + \sum_{i=1}^q \beta_i \sigma_{t-i}^2 
\end{equation}
where $\alpha_i$, with $i = 1, \ldots, p$, and $\beta_i$, with $i = 1, \ldots, q$ are the parameters of the GARCH part of the model, and $\varepsilon_t$ follows a Student's t distribution.

\subsubsection{Google trends (\texttt{Google})}\label{sec:Google}

Since the \texttt{Google} marginal in all locations includes several values equal to zero, we fitted a zero adjusted gamma distribution (ZAGA) using time as explanatory variable (see \cite{rigby2005generalized}). This distribution is a mixture of a discrete value $0$ with probability $\nu$, and a gamma distribution on the positive real line $(0,\infty)$ with probability $(1 - \nu)$.
The probability density function (pdf) of the ZAGA model is given by
\begin{equation}
    f_X(x | \mu, \sigma, \nu) = 
    \begin{cases}
    \nu & \text{if } x=0\\
    (1-\nu) f_{GA}(x | \mu, \sigma) & \text{if } x > 0
    \end{cases}
\end{equation}
for $0 \leq x < \infty$, $0 < \nu < 1$, where $\mu > 0$ is the scale parameter, $\sigma > 0$ is the shape parameter and $f_{GA}(x | \mu, \sigma)$ is the gamma pdf.
We assumed that the parameter $\mu$ of the ZAGA model is related to time, as explanatory variable, through an appropriate link function, with coefficient $\beta$ (for more details, see \cite{rigby2019distributions}).

\subsubsection{Total number of tweets (\texttt{Total\_tweets})}

The best fitting model for the marginal \texttt{Total\_tweets} is the zero adjusted inverse Gaussian distribution (ZAIG), which is similar to the ZAGA model discussed in Section \ref{sec:Google}.
The pdf of the ZAIG model is
\begin{equation}
    f_X(x | \mu, \sigma, \nu) = 
    \begin{cases}
    \nu & \text{if } x=0\\
    (1-\nu) f_{IG}(x | \mu, \sigma) & \text{if } x > 0
    \end{cases}
\end{equation}
for $0 \leq x < \infty$, $0 < \nu < 1$, where $\mu > 0$ is the location parameter, $\sigma > 0$ is the scale parameter and $f_{IG}(x | \mu, \sigma)$ is the inverse Gaussian pdf.
Similarly to the ZAGA model, for the ZAIG model we assumed that the parameter $\mu$ is related to time, as explanatory variable, through an appropriate link function, with coefficient $\beta$  (see \cite{rigby2005generalized, rigby2019distributions}).

\subsubsection{Bing sentiment score (\texttt{Bing})}

The best model for the \texttt{Bing} marginal for all three locations was the ARIMA-GARCH model with Student's t innovations, as illustrated in Eq.\eqref{eq:ARIMA-GARCH}, fitted on the log-transformed data.

Since the residuals of the Dawlish data still showed some structure, they were fitted using a Generalized t distribution (GT), which depends on four parameters controlling location, scale and kurtosis (for more information, see \cite{rigby2005generalized, rigby2019distributions}).

\subsubsection{Afinn sentiment score (\texttt{Afinn})}

The log-transformed \texttt{Afinn} marginal was fitted with an ARIMA-GARCH model with Student's t innovations (see Eq.\eqref{eq:ARIMA-GARCH}).

For Portsmouth, since the residuals still presented some structure, they were fitted using a skew exponential power type 2 distribution (SEP2), which depends on four parameters: the location, scale, skewness and kurtosis. For the implementation of the SEP2 distribution, we used time as explanatory variable for the location parameter.

For Dawlish, the residuals were fitted using a Normal-exponential-Student-t distribution (NET), considering time as explanatory variable. The NET distribution is symmetric and depends on four parameters controlling the location, scale and kurtosis (for more details on the SEP2 and NET distributions, see \cite{rigby2005generalized, rigby2019distributions}).

\vspace{0.5cm}

%%%%%%%%%%%%%%%%%%%%%%%%%%%%%%%%%%%
%%     Vine Copula Model 		%%%
%%%%%%%%%%%%%%%%%%%%%%%%%%%%%%%%%%%

\subsection{Vine Copula Model}\label{sub:vinecopula}

A \textit{vine copula} (or \textit{vine}) represents the pattern of dependence of multivariate data via a cascade of bivariate copulas, allowing us to construct flexible high-dimensional copulas using only bivariate copulas as building blocks. For more details about vine copulas see \cite{czado2019analyzing}.

In order to obtain a vine copula we proceed as follows.
First we factorise the joint distribution $f(x_1, \ldots,x_d)$ of the random vector $\textbf{X}=(X_1, \ldots, X_d)$ as a product of conditional densities
\begin{equation}\label{eq:factor}
f(x_1, \ldots,x_d) = f_d(x_d) \cdot f_{d-1|d}(x_{d-1}|x_d) \cdot \ldots \cdot f_{1|2 \ldots d}(x_1|x_2, \ldots x_d).
\end{equation}
The factorisation in \eqref{eq:factor} is unique up to re-labelling of the variables and it can be expressed in terms of a product of bivariate copulas.
In fact, by Sklar's theorem, the conditional density of $X_{d-1}|X_d$ can be easily written as
\begin{equation}\label{eq:conditional}
f_{d-1|d}(x_{d-1}|x_d) = c_{d-1,d}(F_{d-1}(x_{d-1}), F_d(x_d);\boldsymbol{\theta}_{d-1,d}) \cdot f_{d-1}(x_{d-1}),
\end{equation}
where $c_{d-1,d}$ is a bivariate copula, with parameter vector $\boldsymbol{\theta}_{d-1,d}$. 
Through a straightforward generalisation of Eq.\eqref{eq:conditional}, each term in \eqref{eq:factor} can be decomposed into the appropriate bivariate copula times a conditional marginal density.
More precisely, for a generic element $X_j$ of the vector $\textbf{X}$ we obtain
\begin{equation}\label{eq:condgen}
f_{X_j|\textbf{V}}(x_j|\textbf{v}) = c_{X_J,\nu_{\ell};\textbf{V}_{-\ell}}(F_{X_j|\textbf{V}_{-\ell}}(x_j|\textbf{v}_{-\ell}), F_{\nu_{\ell}|\textbf{V}_{-\ell}}(\nu_{\ell}|\textbf{v}_{-\ell}); \boldsymbol{\theta}_{X_J,\nu_{\ell};\textbf{V}_{-\ell}}) \cdot f_{X_j|\textbf{V}_{-\ell}}(x_j|\textbf{v}_{-\ell}),
\end{equation}
where $\textbf{v}$ is the conditioning vector, $\nu_{\ell}$ is a generic component of $\textbf{v}$, $\textbf{v}_{-\ell}$ is the vector $\textbf{v}$ without the component $\nu_{\ell}$, $F_{X_j|\textbf{v}_{-\ell}}(\cdot|\cdot)$ is the conditional distribution of $X_j$ given $\textbf{v}_{-\ell}$ and $c_{X_J,\nu_{\ell};\textbf{V}_{-\ell}}(\cdot, \cdot)$ is the conditional bivariate copula density, which can typically belong to any family (e.g. Gaussian, Student’s t, Clayton, Gumbel, Frank, Joe, BB1, BB6, BB7, BB8, etc.; for more information on copula families, see \cite{nelsen2007introduction}), with parameter $\boldsymbol{\theta}_{X_J,\nu_{\ell};\textbf{V}_{-\ell}}$. The $d$-dimensional joint multivariate distribution function can hence be expressed as a product of bivariate copulas and marginal distributions by recursively plugging Eq.\eqref{eq:condgen} in Eq.\eqref{eq:factor}.

For example, let us consider a $6$-dimensional distribution. Then, Eq.\eqref{eq:factor} translates to
\begin{equation}\label{eq:factor6}
f(x_1, \ldots,x_6) = f_6(x_6) \cdot f_{5|6}(x_{5}|x_6) \cdot f_{4|5,6}(x_{4}|x_5, x_6) \cdot \ldots \cdot f_{1|2 \ldots 6}(x_1|x_2, \ldots x_6).
\end{equation}
The second factor $f_{5|6}(x_{5}|x_6)$ on the right-hand side of \eqref{eq:factor6} can be easily decomposed into the bivariate copula $c_{5,6}(F_5(x_5), F_6(x_6))$ and marginal density $f_5(x_5)$:
$$
f_{5|6}(x_{5}|x_6) = c_{5,6}(F_5(x_5), F_6(x_6); \boldsymbol{\theta}_{5,6}) \cdot f_5(x_5).
$$
On the other hand, the third factor on the right-hand side of \eqref{eq:factor6} can be decomposed using the \eqref{eq:condgen} as
$$
f_{4|5,6}(x_{4}|x_5, x_6) = c_{4,5;6}(F_{4|6}(x_4|x_6), F_{5|6}(x_5|x_6); \boldsymbol{\theta}_{4,5;6}) \cdot f_{4|6}(x_4|x_6).
$$
Therefore, one of the possible decompositions of the joint density $f(x_1, \ldots,x_6)$ is given by the following expression, which includes the product of marginal densities and copulas, which are all bivariate: 
\begin{align}
f(x_1, \ldots,x_6) =  \prod_{j=1}^6 f_j(x_j) & \cdot c_{1,2} \cdot c_{1,3} \cdot c_{3,4} \cdot c_{1,5} \cdot c_{5,6} \cdot c_{2,3;1} \cdot c_{1,4;3} \cdot c_{3,5;1} \cdot c_{1,6;5} \notag \\
& \cdot  c_{2,4;1,3} \cdot c_{4,5;1,3} \cdot c_{3,6;1,5} \cdot c_{2,5;1,3,4} \cdot c_{4,6;1,3,5} \cdot c_{2,6;1,3,4,5}. \label{eq:pcc}
\end{align}
Eq.\eqref{eq:pcc} is called \textit{pair copula construction}.
Note that in the previous equation the notation has been simplified, setting $c_{a,b} = c_{a,b}(F_a(x_a),F_b(x_b); \boldsymbol{\theta}_{a,b})$.

Two particular types of vines are the Gaussian vine and the Independence vine. The first one is constructed using solely Gaussian bivariate pair-copulas as building blocks, such that each conditional bivariate copula density $c_{X_J,\nu_{\ell};\textbf{V}_{-\ell}}(\cdot, \cdot)$ described in Eq.\eqref{eq:condgen} is a Gaussian copula. The Gaussian vine was adopted in flood risk analysis by \cite{couasnon2018copula}.  The second type is the independence vine, which is constructed using only independence pair-copulas, that are simply given by the product of the marginal distributions of the random variables. In this latter case each conditional bivariate copula density $c_{X_J,\nu_{\ell};\textbf{V}_{-\ell}}(\cdot, \cdot)$ described in Eq.\eqref{eq:condgen} is an Independence copula, implying absence of dependence between the variables.

Pair copula constructions can be represented through a graphical model called \textit{regular vine} (R-vine).
An R-vine $\mathcal{V}(d)$ on $d$ variables is a nested set of trees (connected acyclic graphs) $T_1, \ldots, T_{d-1}$, where the variables are represented by nodes linked by edges, each associated with a certain bivariate copula in the corresponding pair copula construction. The edges of tree $T_k$ are the nodes of tree $T_{k+1}$, $k = 1, \ldots, d-1$. 
Two edges can share a node in tree $T_k$ without the associated nodes in tree $T_{k+1}$ being connected. In an R vine, two edges in $T_k$ which become two nodes in tree $T_{k+1}$, can only share an edge if in tree $T_k$ the edges shared a common node, but they are not necessarily connected by an edge. 
\begin{figure}[h]
    \centering
    \includegraphics[width=0.9\textwidth]{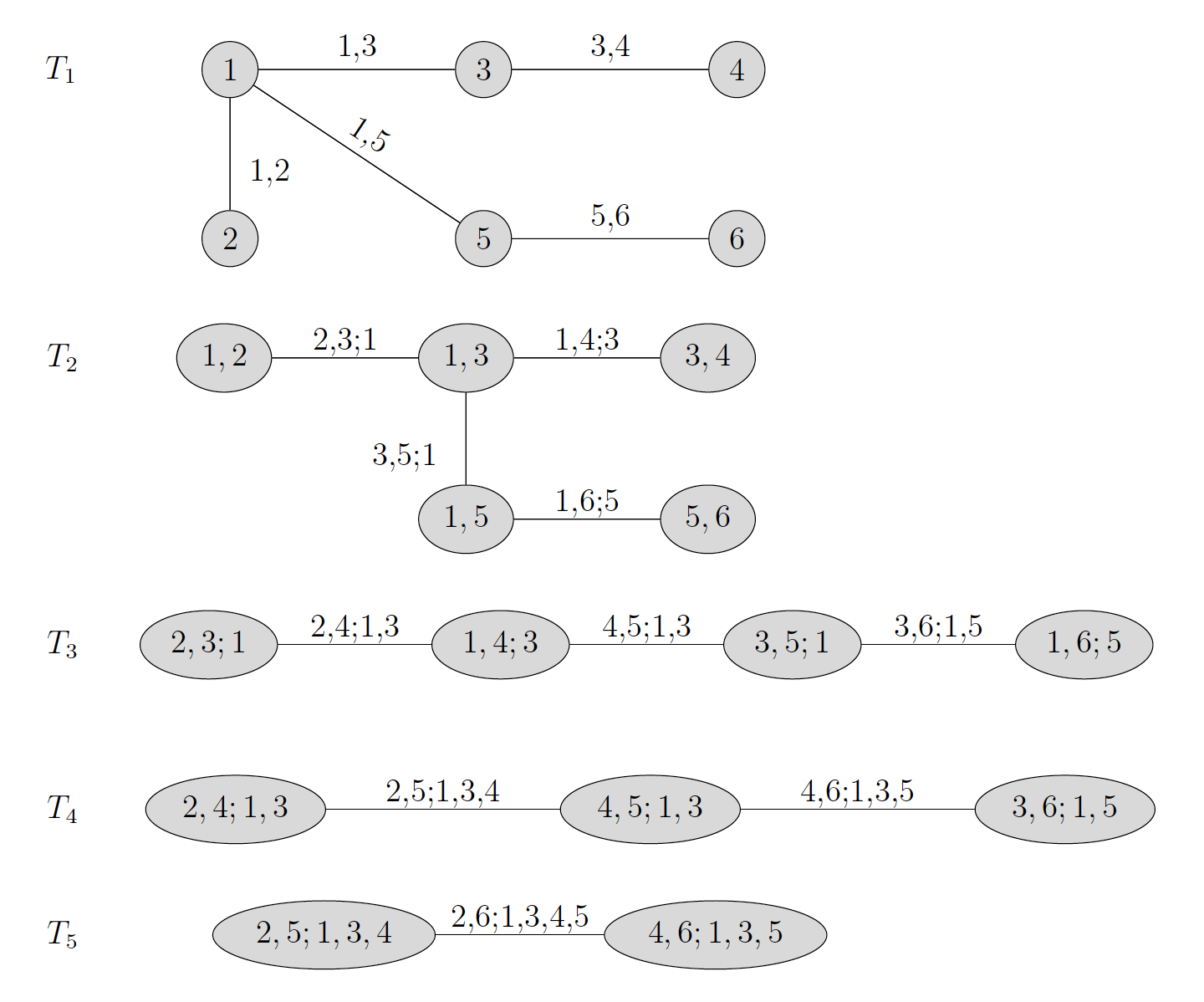}
    \caption{Six-dimensional R-vine graphical representation. \textit{Source: \cite{czado2019analyzing}}}
    \label{fig:6dimVine}
\end{figure}
Figure \ref{fig:6dimVine} shows the 6-dimensional R-vine represented in Eq.\eqref{eq:pcc}. Each edge corresponds to a pair copula density (possibly belonging to different families) and the edge label corresponds to the subscript of the pair copula density, e.g. edge $2,4;1,3$ corresponds to the copula $c_{2,4;1,3}$.

In order to estimate the vine, its structure as well as the copula parameters have to be specified. 
A sequential
approach is generally adopted to select a suitable R-vine decomposition, specifying the first tree and then proceeding similarly for the
following trees. For selecting the structure of each tree, we followed the approach suggested
by \cite{aas2009pair} and developed by \cite{dissmann2013selecting}, using the maximal spanning tree
algorithm. This algorithm defines a tree on all nodes (named spanning tree), which
maximizes the sum of absolute pairwise dependencies, measured, for example, by
Kendall's $\tau$. This specification allows us to capture the strongest dependencies in the first
tree and to obtain a more parsimonious model.
Given the selected tree structure, a copula family for each pair of variables is
identified using the Akaike Information Criterion (AIC), or the Bayesian Information
Criterion (BIC). This choice is typically made amongst a large set of families, comprising elliptical
copulas (Gaussian and Student's t) as well as Archimedean copulas (Clayton, Gumbel, Frank and
Joe), their mixtures (BB1, BB6, BB7 and BB8) and their rotated versions, to cover a large range of possible
dependence structures. For an overview of the different copula families, see \cite{joe1997multivariate} or \cite{nelsen2007introduction}.
The copula parameters $\boldsymbol{\theta}$ for each pair-copula in the vine are estimated using the maximum likelihood (MLE) method, as illustrated by \cite{aas2009pair}. The R-vine estimation procedure is repeated for all the trees, until the R-vine is completely specified.

%%%%%%%%%%%%%%%%%%%%%%%%%%%%%%%%%%%%%%%%%%%%%%%
%%     Out-of-sample predictions 			%%%
%%%%%%%%%%%%%%%%%%%%%%%%%%%%%%%%%%%%%%%%%%%%%%%

\subsection{Out-of-sample predictions}\label{sec:outofsample_pred}

In order to evaluate the suitability of the proposed vine copula model in relation to other methods, we produced one-day-ahead out-of-sample predictions and we compared them to the original data. 
Let $\textbf{X} = \{ \textbf{X}_t; t = 1, \ldots, T \}$ be the $6$-dimensional time series of environmental and social media data. Our aim is to forecast $\textbf{X}_{T+1}$ based on the information available at time $T$. In order to do that, we adopted the forecasting method described by \cite{simard2015forecasting}.
Before fitting the vine, we extracted the residuals from the marginals, as explained in Section \ref{sub:marginals}, and obtained the \textit{u-data}. Next, after fitting the vine, we simulated $M$ realizations from the vine copula.
Hence, we
calculated the predicted values for each simulation, using the inverse cdf and the relevant fitted marginal models.
More precisely, we applied the inverse transformation to the $M$ realizations from the vine copula to obtain the residuals which we then plugged into the marginal models to get the predicted values of the environmental variables (wave height and water level).
Then, we calculated the average prediction for all simulations $\hat{\textbf{X}}_{T+1}^{Avg}$ and use it as the forecast $\textbf{X}_{T+1}$.
The prediction interval of level $(1 - \alpha) \in (0, 1)$ for $\textbf{X}_{T+1}$  was calculated by taking the estimated
quantiles of order $\alpha/2$ and $1 - \alpha/2$ amongst the simulated data. We denote by $\hat{\textbf{X}}_{T+1}^l$
and $\hat{\textbf{X}}_{T+1}^u$
the lower and upper values of the prediction intervals.

In order to compare and contrast the accuracy of predictions for different models, we made use of four indicators: the mean squared error (MSE) to evaluate point forecasts; the mean interval score (MIS), proposed by \cite{gneiting2007strictly}, to assess the accuracy of the prediction intervals; the Normalized Nash-Sutcliffe model efficiency (NNSE) coefficient, proposed by \cite{nash1970river}
to appraise hydrological  models; and the Distance Correlation, proposed by \cite{szekely2007measuring}, to determine the association between observed and predicted data.
The MSE for each variable $j = 1,\ldots, d$ was calculated as follows
$$
\mbox{MSE}_j = \frac{1}{S} \sum_{t=T+1}^{T+S} 
(x_{t,j} - \hat{x}_{t,j})^2
$$
where $x_{t,j}$ is the observed value for each variable at each time point $t$, $\hat{x}_{t,j}$ is the corresponding predicted value, $T+1$ denotes the first predicted date, while $T+S$ indicates the last predicted date. 
The 95\% MIS for each variable, at level $\alpha=0.05$, was computed as
$$
\mbox{MIS}_j = \frac{1}{S} \sum_{t=T+1}^{T+S} \left[ (\hat{x}_{t,j}^u - \hat{x}_{t,j}^l) + \frac{2}{\alpha} (\hat{x}_{t,j}^l - x_{t,j}) \mathbbm{1}(x_{t,j} < \hat{x}_{t,j}^l) + \frac{2}{\alpha} (x_{t,j} - \hat{x}_{t,j}^u) \mathbbm{1}(x_{t,j} > \hat{x}_{t,j}^u) \right]
$$
where $\hat{x}_{t,j}^l$ and $\hat{x}_{t,j}^u$ denote, respectively, the lower and upper limits of the prediction intervals for each variable at each time point, and $\mathbbm{1}(\cdot)$ is the indicator function.
The NNSE coefficient was calculated as 
$$
\mbox{NNSE}_j = \frac{1}{2-\mbox{NSE}_j}
$$
with
$$ 
\mbox{NSE}_j = 1 -\frac { \sum_{t=T+1}^{T+S} { \left( \hat{x}_{t,j} - x_{t,j} \right)^2 } } { \sum_{t=T+1}^{T+S} { \left( x_{t,j} - \bar{x}_{j} \right)^2 } } 
$$
where $\bar{x}_j$ is the mean of the observed values for each variable. The NSE is a normalized statistic that determines the relative magnitude of the residual variance (``noise'') compared to the measured data variance (``information''). The Distance Correlation takes the form
$$
\mbox{DC}_j = \mbox{dCor}(X_j,\hat{X}_j) = \frac{\mbox{dCov}(X_j,\hat{X}_j)}{\sqrt{\mbox{dVar}(X_j)\mbox{dVar}(\hat{X}_j)}}
$$
where $X_j$ is the $j$-th observed variable, $\hat{X}_j$ is the corresponding $j$-th predicted variable, $\mbox{dCor}(X_j,\hat{X}_j)$ is the distance covariance and $\mbox{dVar}(X_j)$ and $\mbox{dVar}(\hat{X}_j)$ are the distance standard deviations, obtained replacing the signed distances between the variables with centred Euclidean distances. The DC is a distance-based correlation that can detect both linear and non-linear relationships between variables.

%%%%%%%%%%%%%%%%%%%%%%%%%%%%%%%%%%%
%%     Result Analysis 			%%%
%%%%%%%%%%%%%%%%%%%%%%%%%%%%%%%%%%%

\section{Result Analysis and Discussion}\label{sec:analysis}

We now present the results of the analysis of the remotely-sensed and online flood data for the three locations under consideration.

\subsection{Twitter Wordclouds}

First, we analysed the information gathered on Twitter,
cleaning and stemming the tweets and
producing wordclouds for each location.

\begin{figure}[htbp]
    \centering
    \includegraphics[width=0.4\textwidth]{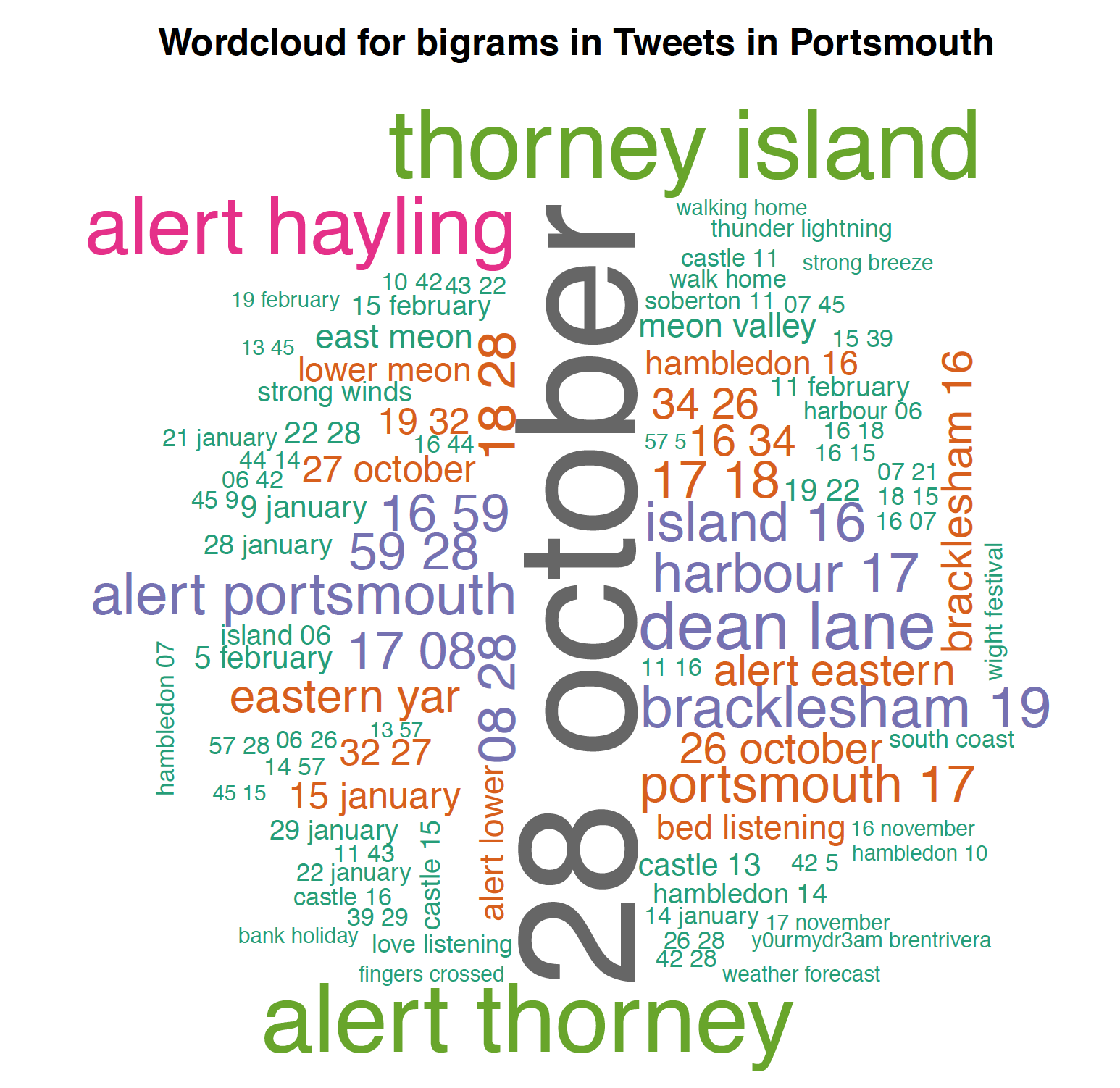} \\
    \includegraphics[width=0.4\textwidth]{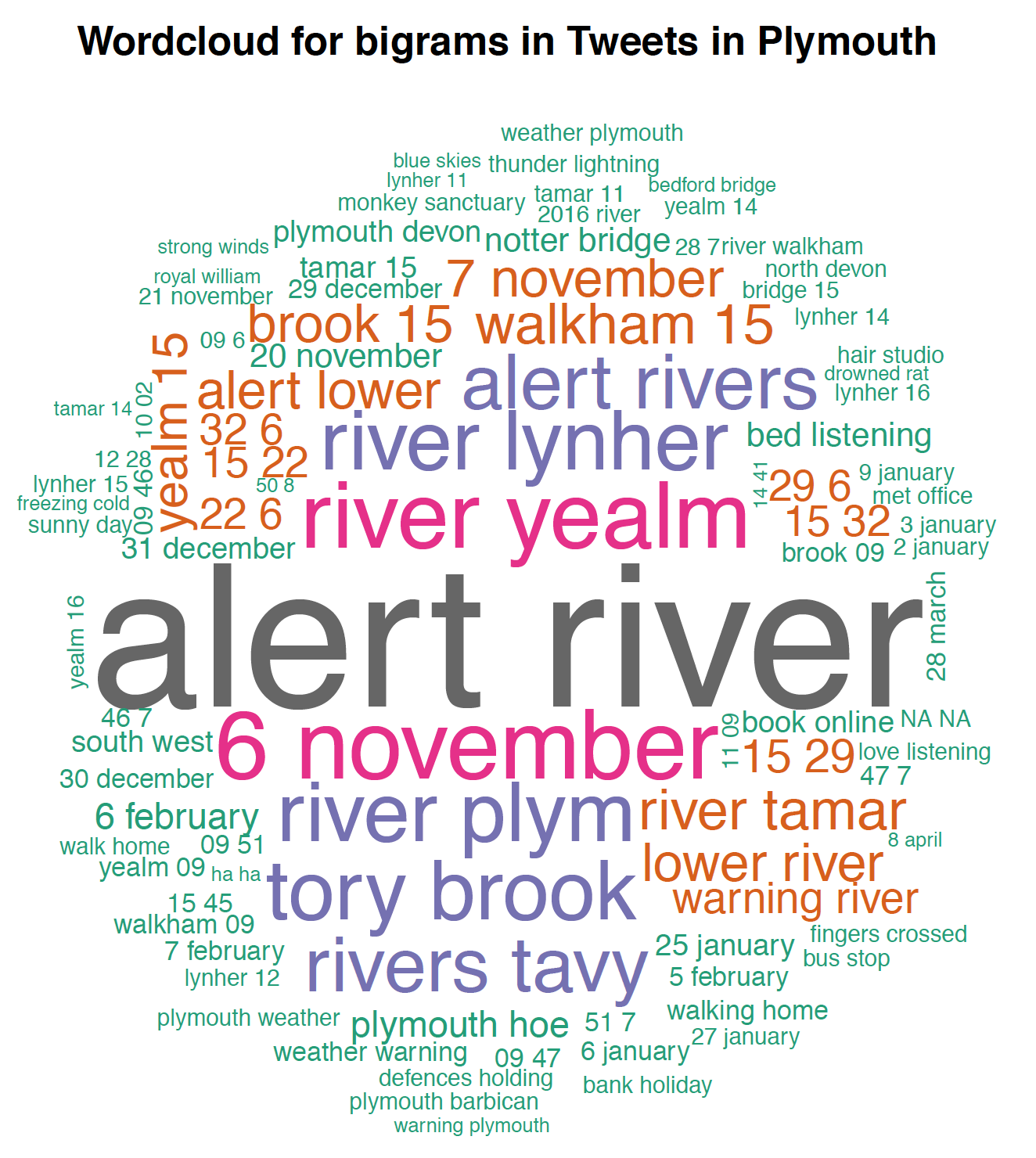} \\
    \includegraphics[width=0.4\textwidth]{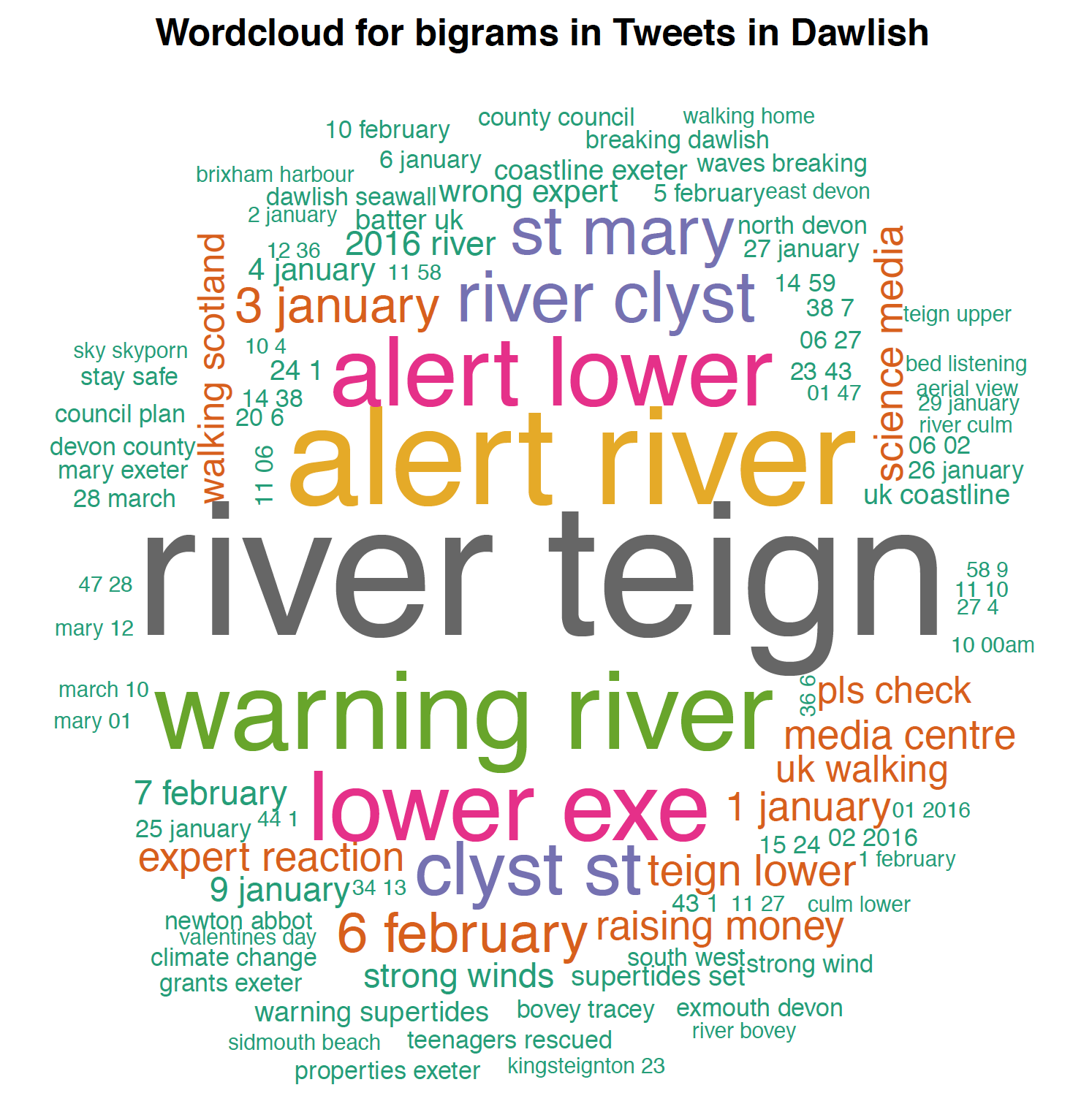}
    \caption{Wordcloud of paired words in tweets from Portsmouth (top panel), Plymouth (middle panel) and Dawlish (bottom panel).}
    \label{fig:bigrams}
\end{figure}

Figure \ref{fig:bigrams} displays the wordclouds of paired words obtained by pairing the the most common combinations of words appearing in the collected tweets.  
The top, middle and bottom panels show the wordclouds of Portsmouth, Plymouth and Dawlish tweets, respectively.
The most frequent pairs of words refer to dates indicating storm and flood events (e.g. 28 October, 3 January), names of places affected by storms (e.g. Thorney Island, St Mary) and names of rivers (e.g. river Yealm, river Teign).

\subsection{Marginals Estimation}

Table \ref{tab:marg_par} lists the parameter estimates, obtained via the MLE method, of the best fitting models for the marginals, as described in Section \ref{sub:marginals}, for Portsmouth (top panels), Plymouth (middle panels) and Dawlish (bottom panels). Standard errors are in brackets.\footnote{Please, note that, due to lack of space, the Table does not include the estimates of the GT, SEP2 and NET models fitted to the residuals of the \texttt{Bing} and \texttt{Afinn} marginals.}  

\begin{figure}[htbp]
    \centering
    \includegraphics[width=0.6\textwidth]{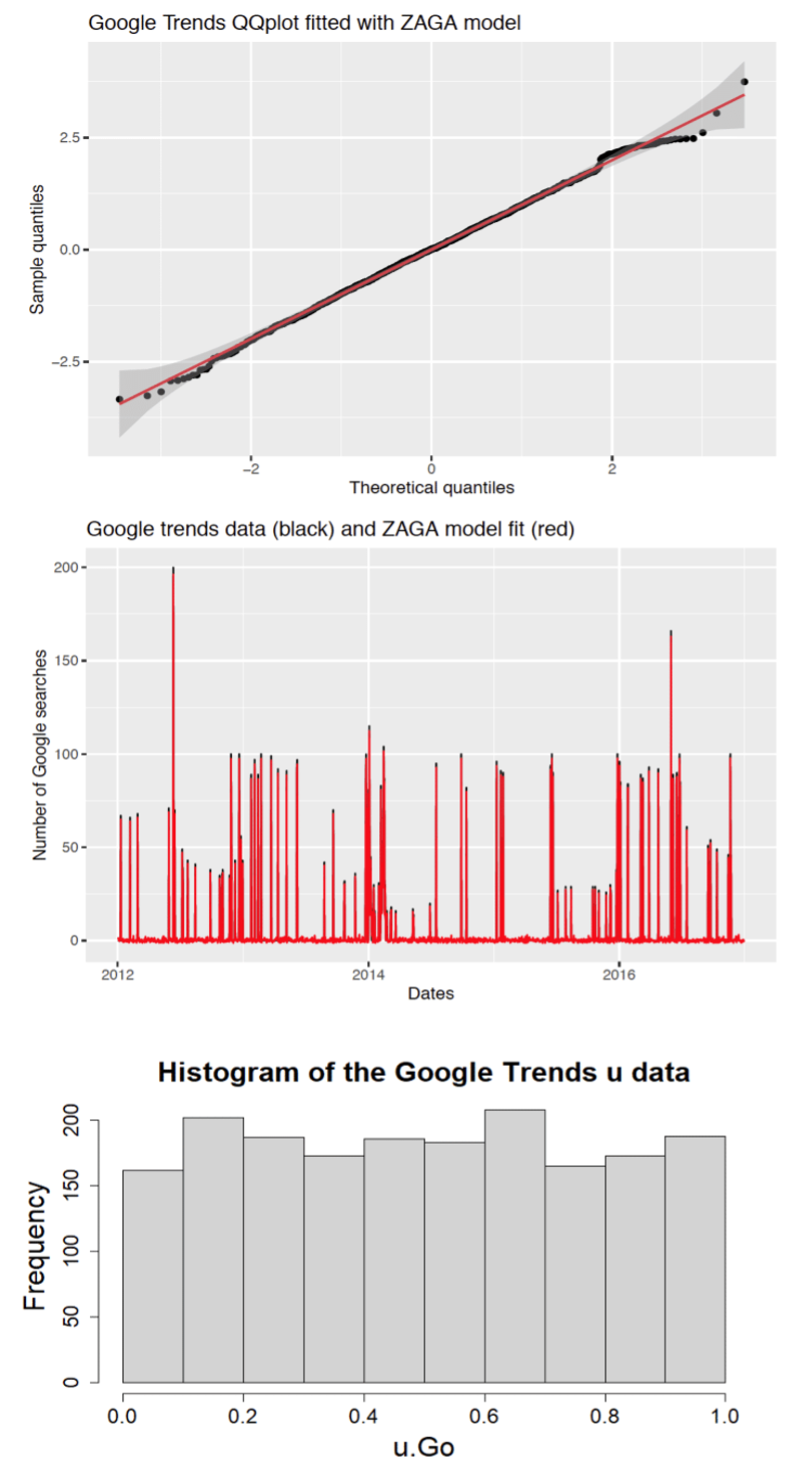}
    \caption{Plot illustrating the fit of the residuals for the \texttt{Google} marginal for Portsmouth. Top panel: QQ-plot comparing the Gaussian theoretical quantiles with sample quantiles. Middle panel: observed time series (black line) and in-sample predictions obtained from the fitted ZAGA model (red line). Bottom panel: Histogram of the resulting \textit{u-data}.}
    \label{fig:qqplot_insample}
\end{figure}

As an example, Figure \ref{fig:qqplot_insample} shows the fit of the residuals for the \texttt{Google} trends marginal for Portsmouth. The other plots for all marginals related to all three locations exhibit a similar behaviour. The top panel displays the QQ-plot comparing the Gaussian theoretical quantiles with the sample quantiles, the middle panel illustrates the observations (black line) and in-sample predictions obtained from the fitted ZAGA model (red line), while the bottom panel shows the histogram of the resulting u-data. The plots clearly show an excellent fit of the ZAGA model to the marginal, as demonstrated by the points in the QQ-plot aligning almost perfectly to the main diagonal, the in-sample predictions overlapping the observed data and the shape of the \textit{u-data} histogram displaying a uniform pattern.

\begin{landscape}
\begin{table}[htbp]
  \centering
  \caption{Parameter estimates of the marginals for each location: Portsmouth (top panels), Plymouth (middle panels) and Dawlish (bottom panels). Standard errors are in brackets.}
  \begin{scriptsize}
    \begin{tabular}{|rr|lr|lr|lr|lr|rr|}
    \hline
    \multicolumn{12}{|c|}{\textbf{Marginals}} \\
    \hline
    \multicolumn{12}{|c|}{Portsmouth} \\
    \hline
    \multicolumn{2}{|c|}{\texttt{Hs}} & \multicolumn{2}{c|}{\texttt{WL}} & \multicolumn{2}{c|}{\texttt{Google}} & \multicolumn{2}{c|}{\texttt{Total\_tweets}} & \multicolumn{2}{c|}{\texttt{Bing}} & \multicolumn{2}{c|}{\texttt{Afinn}} \\
    \hline
    \multicolumn{2}{|c|}{ARIMA(3,0,2)} & \multicolumn{2}{c|}{ARIMA(1,0,1)-GARCH(1,1)} & \multicolumn{2}{c|}{ZAGA} & \multicolumn{2}{c|}{ZAIG} & \multicolumn{2}{c|}{ARIMA(1,d,1)-GARCH(3,1)} & \multicolumn{2}{c|}{ARIMA(1,d,0)-GARCH(3,1)	} \\
    \hline
    \multicolumn{1}{|l}{$a$} & -0.1146 (0.0730) & $a$     & 0.1373    (0.0066) & $\mu$ & 2.2579 (1.5760) & $\mu$ & 8.8258 (1.3940) & a     & 1.0000 (0.00009) & \multicolumn{1}{l}{a} & 1.0000 (0.00009) \\
    \multicolumn{1}{|l}{$\phi_1$} & 0.9810 (0.2135) & $\phi_1$ & 0.7679   ( 0.0221) & $\beta$ & 0.0001 (0.00009) & $\beta$ & -0.0004 (0.00008) & $\phi_1$ & 0.8707 (0.0172) & \multicolumn{1}{l}{$\phi_1$} & 0.3832 (0.0435) \\
    \multicolumn{1}{|l}{$\phi_2$} & 0.2891 (0.3474) & $\theta_1$ & -0.1174    (0.0351) & $\sigma$ & -0.5203 (0.0581) & $\sigma$ & -0.6741 (0.0197) & $\theta_1$ & -0.7335 (0.0351) & \multicolumn{1}{l}{d} & 0.2836 (0.1092) \\
    \multicolumn{1}{|l}{$\phi_3$} & -0.2836 (0.1382) & $\omega$ & 0.0001    (0.00006) & $\nu$ & 2.5440 (0.0901) & $\nu$ & -0.8606 (0.0512) & d     & 0.3382 (0.0363) & \multicolumn{1}{l}{$\omega$} & 0.0000 (0.0000) \\
    \multicolumn{1}{|l}{$\theta_1$} & -0.3112 (0.2060) & $\alpha_1$ & 0.0651    (0.0162) &       &       &       &       & $\omega$ & 0.0000 (0.0000) & \multicolumn{1}{l}{$\alpha_1$} & 0.0167 (0.0038) \\
    \multicolumn{1}{|l}{$\theta_2$} & -0.5775 (0.1979) & $\beta_1$ & 0.9214    (0.0196) &       &       &       &       & $\alpha_1$ & 0.0167 (0.0037) & \multicolumn{1}{l}{$\alpha_2$} & 0.0167 (0.0097) \\
          &       & $\sigma$ & 5.9142    (0.7878) &       &       &       &       & $\alpha_2$ & 0.0167 (0.0086) & \multicolumn{1}{l}{$\alpha_3$} & 0.0167 (0.0075) \\
          &       &       &       &       &       &       &       & $\alpha_3$ & 0.0167 (0.0072) & \multicolumn{1}{l}{$\beta_1$} & 0.9000 (0.0111) \\
          &       &       &       &       &       &       &       & $\beta_1$ & 0.9000 (0.0096) & \multicolumn{1}{l}{$\sigma$} & 3.9999 (0.2633) \\
          &       &       &       &       &       &       &       & $\sigma$ & 4.0000 (0.2617) &       &  \\
    \hline
    \multicolumn{12}{|c|}{Plymouth} \\
    \hline
    \multicolumn{2}{|c|}{ARIMA(1,0,2)} & \multicolumn{2}{c|}{ARIMA(4,1,1)} & \multicolumn{2}{c|}{ZAGA} & \multicolumn{2}{c|}{ZAIG} & \multicolumn{2}{c|}{ARIMA(1,d,2)-GARCH(2,1)} & \multicolumn{2}{c|}{ARIMA(1,d,1)-GARCH(2,1)	} \\
    \hline
    \multicolumn{1}{|l}{a} & 0.0048 (0.0473) & $\phi_1$ & 0.7204 (0.0240) & $\mu$ & 4.927 (1.403) & $\mu$ & 4.826 (1.371) & a     & 1.0000 (0.00005) & \multicolumn{1}{l}{a} & 1.0000 (0.00009) \\
    \multicolumn{1}{|l}{$\phi_1$} & 0.8366 (0.0248) & $\phi_2$ & 0.0582 (0.0289) & $\beta$ & -0.00005 (0.00008) & $\beta$ & -0.0002 (0.00008) & $\phi_1$ & 0.9779 (0.0007) & \multicolumn{1}{l}{$\phi_1$} & 0.3229 (0.0882) \\
    \multicolumn{1}{|l}{$\theta_1$} & -0.1016 (0.0368) & $\phi_3$ & 0.0035 (0.0289) & $\sigma$ & -0.3348 (0.0395) & $\sigma$ & -0.6058 (0.0219) & $\theta_1$ & -0.8854 (0.0115) & \multicolumn{1}{l}{$\theta_1$} & -0.1301 (0.0968) \\
    \multicolumn{1}{|l}{$\theta_2$} & -0.1598 (0.0337) & $\phi_4$ & 0.0112 (0.0239) & $\nu$ & 1.7263 (0.0653) & $\nu$ & -0.2788 (0.0473) & $\theta_2$ & -0.0553 (0.0115) & \multicolumn{1}{l}{d} & 0.3111 (0.0683) \\
          &       & $\theta_1$ & -0.9919 (0.0055) &       &       &       &       & d     & 0.3076 (0.0490) & \multicolumn{1}{l}{$\omega$} & 0.0000 (0.0000) \\
          &       &       &       &       &       &       &       & $\omega$ & 0.0000 (0.0000) & \multicolumn{1}{l}{$\alpha_1$} & 0.0250 (0.0062) \\
          &       &       &       &       &       &       &       & $\alpha_1$ & 0.0250 (0.0068) & \multicolumn{1}{l}{$\alpha_2$} & 0.0250 (0.0057) \\
          &       &       &       &       &       &       &       & $\alpha_2$ & 0.0250 (0.0051) & \multicolumn{1}{l}{$\beta_1$} & 0.9000 (0.0137) \\
          &       &       &       &       &       &       &       & $\beta_1$ & 0.9000 (0.0069) & \multicolumn{1}{l}{$\sigma$} & 4.0000 (0.5849) \\
          &       &       &       &       &       &       &       & $\sigma$ & 4.0000 (0.3481) &       &  \\
    \hline
    \multicolumn{12}{|c|}{Dawlish} \\
    \hline
    \multicolumn{2}{|c|}{ARIMA(2,0,3)} & \multicolumn{2}{c|}{ARIMA(1,0,1)-GARCH(1,1)	} & \multicolumn{2}{c|}{ZAGA} & \multicolumn{2}{c|}{ZAIG} & \multicolumn{2}{c|}{ARIMA(2,d,1)-GARCH(1,1)} & \multicolumn{2}{c|}{ARIMA(1,d,1)-GARCH(1,1)	} \\
    \hline
    \multicolumn{1}{|l}{$a$} & -0.3644 (0.0763) & $a$    & 0.1131 (0.0078) & $\mu$ & 0.3969 (2.2670) & $\mu$ & -0.2273 (1.910) & a     & 1.0000 (0.00008) & \multicolumn{1}{l}{a} & 0.9999    (0.00009) \\
    \multicolumn{1}{|l}{$\phi_1$} & 1.5597 (0.1241) & $\phi_1$ & 0.8011 (0.0180) & $\beta$ & 0.0002 (0.0001) & $\beta$ & 0.0015 (0.0001) & $\phi_1$ & -0.2924 (0.0315) & \multicolumn{1}{l}{$\phi_1$} & 0.3049   (0.0268) \\
    \multicolumn{1}{|l}{$\phi_2$} & -0.5674 (0.1208) & $\theta_1$ & -0.0023 (0.0297) & $\sigma$ & -0.0928 (0.0463) & $\sigma$ & -0.6471 (0.0399) & $\phi_2$ & 0.2188 (0.0374) & \multicolumn{1}{l}{$\theta_1$} & 0.0760    (0.0225) \\
    \multicolumn{1}{|l}{$\theta_1$} & -0.9812 (0.1285) & $\omega$ & 0.00008 (0.00004) & $\nu$ & 2.1713 (0.0772) & $\nu$ & 1.5725 (0.0620) & $\theta_1$ & 0.7206 (0.0459) & \multicolumn{1}{l}{d} & 0.3967    (0.0146) \\
    \multicolumn{1}{|l}{$\theta_2$} & -0.0680 (0.0586) & $\alpha_1$ & 0.0496 (0.0102) &       &       &       &       & d     & 0.3139 (0.0028) & \multicolumn{1}{l}{$\omega$} & 0.0000 (0.0000) \\
    \multicolumn{1}{|l}{$\theta_3$} & 0.0998 (0.0627) & $\beta_1$ & 0.9394 (0.0116) &       &       &       &       & $\omega$ & 0.0000 (0.0000) & \multicolumn{1}{l}{$\alpha_1$} & 0.0500   (0.0024) \\
          &       & $\sigma$ & 5.4261 (0.6958) &       &       &       &       & $\alpha_1$ & 0.0500 (0.0023) & \multicolumn{1}{l}{$\beta_1$} & 0.8999    (0.0019) \\
          &       &       &       &       &       &       &       & $\beta_1$ & 0.8999 (0.0022) & \multicolumn{1}{l}{$\sigma$} & 3.9997    (0.1728) \\
          &       &       &       &       &       &       &       & $\sigma$ & 3.999 (0.1399) &       &  \\
    \hline
    \end{tabular}%
    \end{scriptsize}
  \label{tab:marg_par}%
\end{table}%
\end{landscape}

\subsection{Vine Estimation}\label{sec:vine}

Once the marginals were estimated, we derived the corresponding \textit{u-data} from the residuals, as illustrated in Section \ref{sub:marginals}. Then, we carried out fitting and model selection for the vine copula for each location using the R package \texttt{rvinecopulib} \citep{rvinecopulib}.

\begin{figure}[htbp]
    \centering
    \includegraphics[width=1\textwidth]{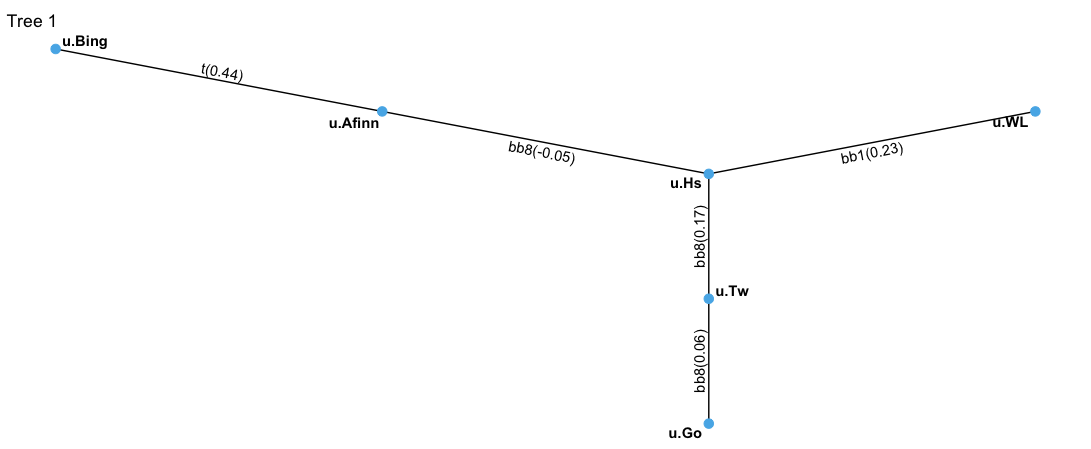} \\
    \vspace{1cm}
    \includegraphics[width=1\textwidth]{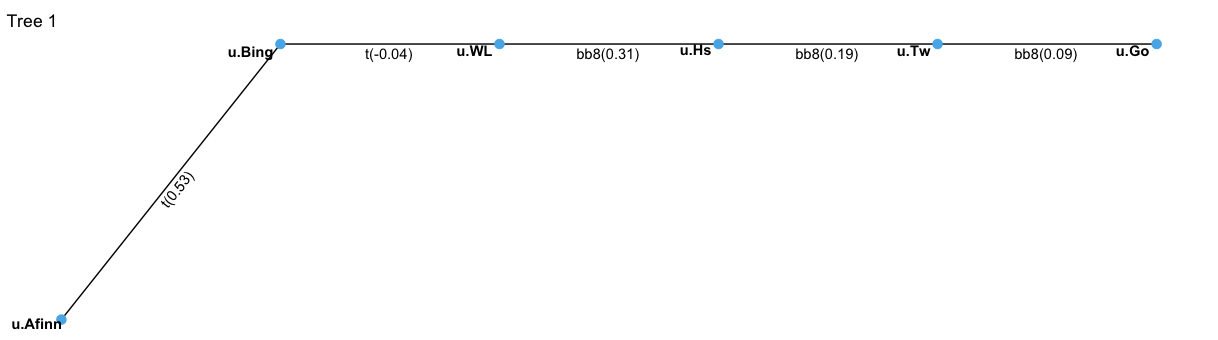} \\
    \vspace{1cm}
    \includegraphics[width=1\textwidth]{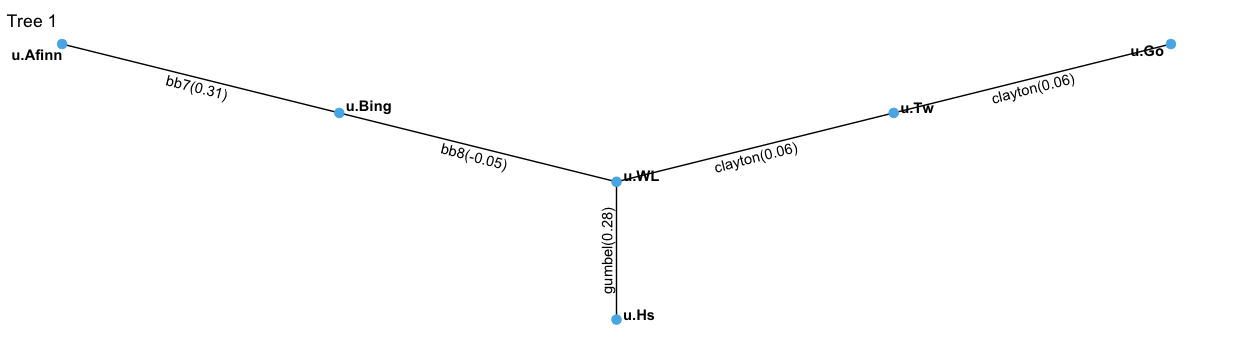}
    \caption{First trees of the vine copulas estimated for Portsmouth (top panel), Plymouth (middle panel) and Dawlish (bottom panel).}
    \label{fig:vine_plots}
\end{figure}

Figure \ref{fig:vine_plots} displays the first trees of the vine copulas estimated for Portsmouth (top panel), Plymouth (middle panel) and Dawlish (bottom panel).
The nodes are denoted with blue dots, with the names of the margins reported in boldface\footnote{Please, note that \texttt{Total\_tweets} is denoted with \texttt{Tw} in the plots.}. On each edge, the plots show the name of the selected pair copula family and the estimated copula parameter expressed as Kendall's $\tau$.
In order to estimate the vines, we adopted the Kendall's $\tau$ criterion for tree selection, the AIC for the copula families selection and the MLE method for estimating the pair copula parameters.
As it is clear from Figure \ref{fig:vine_plots}, the vines for the three different locations exhibit a very similar structure, with the environmental variables \texttt{Hs} and \texttt{WL} playing a central role and linking to the social media variables. The sentiment scores \texttt{Bing} and \texttt{Afinn} are directly associated. Likewise, \texttt{Total\_tweets} and \texttt{Google} are contiguouly related. 
The symmetric Gaussian copula, which is often employed in traditional multivariate modelling, was not identified as the best fitting copula for neither of the locations.
On the contrary, the selected copula families
include the Student's t copula, which is able to model strong tail dependence, Archimedean copulas such as the Clayton and Gumbel, that are able to capture asymmetric dependence, and mixture copulas such as the BB1 (Clayton-Gumbel) and BB8 (Joe-Frank), that can accommodate various dependence shapes. 
Most of the associations between the variables are positive. The strongest associations are between the \texttt{Bing} and \texttt{Afinn} sentiment scores and between the environmental variables \texttt{Hs} and \texttt{WL}. Also, \texttt{Hs} and \texttt{Total\_tweets} are mildly associated.

\subsection{Out-of-sample prediction results}

In this Section we constructed out-of-sample predictions using the proposed vine methodology, which integrates environmental and social media variables. We then compared the predictions obtained with our methodology with those yielded using two traditional approaches. The former is based on vines built exclusively using Gaussian pair copulas, which are the most common in applications, but are restricted to dependence symmetry and absence of tail dependence. The latter approach assumes independence among the six time series under consideration and therefore calculates predictions ignoring any association between environmental and online information.  

Out-of-sample predictions based on the proposed model were constructed as illustrated in Section \ref{sec:outofsample_pred}, considering the vine copula estimated as explained in Section \ref{sec:vine} until the \nth{15} February 2016 and using it to predict the period between the \nth{16} February 2016 and the \nth{31} December 2016. 

%%%%%%%%%%%%%%%%%%%%%%%%%%%
%%%%%%%%%%%% QUIIIII

Tables \ref{tab:mse} and \ref{tab:mis} list the MSE and MIS values calculated for Portsmouth, Plymouth and Dawlish, in the top, middle and bottom panel, respectively, for each variable. The second columns show the results assuming independence among variables, the third columns show the results assuming all Gaussian pair-copulas, and the fourth columns show the vine copula results. The MSEs and MISs of the best performing approaches for each variable are highlighted in boldface.
From Tables \ref{tab:mse} and \ref{tab:mis}, we notice that the vine copula approach outperforms the other two approaches in the majority of the cases. Comparing the three different locations, in Plymouth the vine copula exceeds the performance of the other two approaches for most of the variables, whereas the independence approach is never selected. In Portsmouth the Gaussian vine method achieves generally the best results, with the independence approach only selected in a few cases. In Dawlish, the vine and Gaussian copula methods are preferred for several variables, although the independence approach is selected in a few cases. This might be due to the lack of social media information for Dawlish, compared to the other two locations, as shown in Figure 3, making it difficult to define associations between online and environmental data and to leverage data integration for predicting purposes.

The variables \texttt{Hs} and \texttt{WL} are generally better predicted by the vine method, as opposed to the independence approach, which assumes no dependence between any of the variables involved in the model. Hence, the independence approach indicates the absence of any association between the environmental and the social media variables, implying the lack of contribution of online-generated information in predicting the flood variables. On the contrary, the vine approach assumes the presence of a dependence structure between the variables and, in particular, between the environmental and social media insights. Therefore, the better performance of the vine compared to the independence model demonstrates usefulness of social media information in forecasting environmental variables. 

The prediction of online-generated information also benefits from data integration. Google trends are more accurately forecasted by the vine copula method, or by the Gaussian approach in the Portsmouth case, rather than by the independent approach. The prediction of Total tweets achieves generally better results with the vine copula method for Plymouth data and with the Gaussian method for Portsmouth data, while the independence approach is typically selected for Dawlish data, due to the lack of information for this location, as explained above.

\begin{table}[htbp]
  \centering
  \caption{MSEs calculated for Portsmouth (top panel), Plymouth (middle panel) and Dawlish (bottom panel) for each variable. The figures show the results assuming independence among variables (second column), all Gaussian pair-copulas (third column) and vine copula (fourth column). The MSEs of the best performing approaches for each variable are in boldface.}
{\begin{tabular}{|c |c |c|c|}
\hline
\multicolumn{4}{|c|}{\textbf{MSE Portsmouth}} \\
\hline
Variable & Independent & Gaussian & Vine Copula \\
\hline
\texttt{Hs} & 0.2693 &	\textbf{0.2603} &	 0.2639 \\
\texttt{WL} & \textbf{0.0301}	& 0.0327 &	0.0325  \\
\texttt{Google} & 404.4304 &	\textbf{403.9977} &	404.4147 \\ 
\texttt{Total\_Tweets} & 6.7351 &	\textbf{6.6994} &	6.7829 \\
\texttt{Bing} &  \textbf{2.6572 $\times$ 10$^{-11}$}  &	2.6634 $\times$  10$^{-11}$ & 2.6624 $\times$  10$^{-11}$ \\
\texttt{Afinn} & 1.3823 $\times$ 10$^{-10}$ &	\textbf{1.3745 $\times$ 10$^{-10}$} &	1.3767 $\times$ 10$^{-10}$ \\
\hline
\hline
\multicolumn{4}{|c|}{\textbf{MSE Plymouth}} \\
\hline
Variable & Independent & Gaussian & Vine Copula \\
\hline
\texttt{Hs} & 0.3646 &	0.3647 &	\textbf{0.358} \\
\texttt{WL} &  0.0274 &	0.0278 &	\textbf{0.0261}  \\
\texttt{Google} &  2874.761 &	2875.053 &	\textbf{2873.466} \\ 
\texttt{Total\_Tweets} & 14.2388 &	14.1698 &	\textbf{14.1569} \\
\texttt{Bing} & 2.6834 $\times$ 10$^{-11}$	& \textbf{2.6282 $\times$ 10$^{-11}$} &	2.6829 $\times$ 10$^{-11}$ \\
\texttt{Afinn} & 1.2103 $\times$ 10$^{-10}$ &	1.2035 $\times$ 10$^{-10}$ &	\textbf{1.2028 $\times$ 10$^{-10}$} \\
\hline
\hline
\multicolumn{4}{|c|}{\textbf{MSE Dawlish}} \\
\hline
Variable & Independent & Gaussian & Vine Copula \\
\hline
\texttt{Hs}  & 0.2857 &	\textbf{0.2864} &	0.2915\\
\texttt{WL}  & \textbf{0.0267} & 0.0295 &	0.0285 \\
\texttt{Google} &  4612.772 &	4613.572 &	\textbf{4612.738} \\ 
\texttt{Total\_Tweets} & \textbf{609.9969} &	610.042	& 610.3111 \\
\texttt{Bing} & 5.7304 $\times$ 10$^{-9}$	& \textbf{5.6124 $\times$ 10$^{-9}$} &	5.6264 $\times$ 10$^{-9}$ \\
\texttt{Afinn} & 6.1873 $\times$ 10$^{-9}$ &	6.1670 $\times$ 10$^{-9}$ &	\textbf{6.1208 $\times$ 10$^{-9}$} \\
\hline
\end{tabular}}
\label{tab:mse}%
\end{table}

\begin{table}[htbp]
  \centering
  \caption{MISs calculated for Portsmouth (top panel), Plymouth (middle panel) and Dawlish (bottom panel) for each variable. The figures show the results assuming independence among variables (second column), all Gaussian pair-copulas (third column) and vine copula (fourth column). The MISs of the best performing approaches for each variable are in boldface}
{\begin{tabular}{|c |c |c|c|}
\hline
\multicolumn{4}{|c|}{\textbf{MIS Portsmouth}} \\
\hline
Variable & Independent & Gaussian & Vine Copula \\
\hline
\texttt{Hs} & 0.4193 &	0.4158 &	\textbf{0.1199} \\
\texttt{WL} & \textbf{0.0431} &	0.0436 &	0.0465  \\
\texttt{Google} & 6.1179 &	\textbf{6.1151} & 6.1169 \\ 
\texttt{Total\_Tweets} & 0.6366 &	\textbf{0.6316} &	0.6356 \\
\texttt{Bing} & 1.2021 $\times$ 10$^{-6}$	& \textbf{1.1982 $\times$ 10$^{-6}$} &	1.2039 $\times$ 10$^{-6}$ \\
\texttt{Afinn} & 3.175 $\times$ 10$^{-6}$ &	3.1668 $\times$ 10$^{-6}$ &	\textbf{3.1644 $\times$ 10$^{-6}$} \\
\hline
\hline
\multicolumn{4}{|c|}{\textbf{MIS Plymouth}} \\
\hline
Variable & Independent & Gaussian & Vine Copula \\
\hline
\texttt{Hs} & 0.1554 &	0.1533 &	\textbf{0.1518} \\
\texttt{WL} &  0.0384 &	0.0365 &	\textbf{0.0361} \\
\texttt{Google} &  10.8789	 & 10.879 &	\textbf{10.8759} \\ 
\texttt{Total\_Tweets} & 0.7849 &	0.7845 &	\textbf{0.7833} \\
\texttt{Bing} & 1.2178 $\times$ 10$^{-6}$	& \textbf{1.2043 $\times$ 10$^{-6}$} & 	1.2175 $\times$ 10$^{-6}$ \\
\texttt{Afinn} & 3.0799 $\times$ 10$^{-6}$ & \textbf{3.0693 $\times$ 10$^{-6}$} &	3.0704 $\times$ 10$^{-6}$ \\
\hline
\hline
\multicolumn{4}{|c|}{\textbf{MIS Dawlish}} \\
\hline
Variable & Independent & Gaussian & Vine Copula \\
\hline
\texttt{Hs}  & 0.4177 &	\textbf{0.4116} & 	0.4116 \\
\texttt{WL}  & \textbf{0.0383}	& 0.0388 &	0.0431 \\
\texttt{Google} &  13.5782 & 13.5794 &	\textbf{13.5782} \\ 
\texttt{Total\_Tweets} & \textbf{7.0887} &	7.0889 &	7.0913 \\
\texttt{Bing} & 1.7472 $\times$ 10$^{-5}$ &	\textbf{1.7184 $\times$ 10$^{-5}$} &	1.7284 $\times$ 10$^{-5}$ \\
\texttt{Afinn} & 1.8396 $\times$ 10$^{-5}$ &	1.8385 $\times$ 10$^{-5}$ &	1.8301 $\times$ 10$^{-5}$ \\
\hline
\end{tabular}}
\label{tab:mis}%
\end{table}

Comparing the sentiment scores, we notice that the vine copula approach is generally preferred with \texttt{Afinn}, while the Gaussian method is typically selected with \texttt{Bing}. This is probably due to the fact that the \texttt{Afinn} lexicon is more sophisticated than \texttt{Bing}, since it scores words into several positive and negative categories, and hence provides more information. 

\begin{figure}[htbp]
    \centering
    \includegraphics[width=0.45\textwidth]{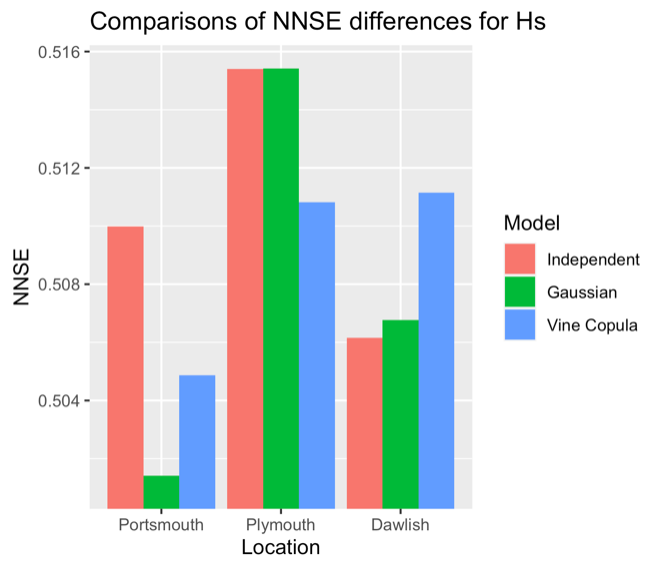} 
    \includegraphics[width=0.45\textwidth]{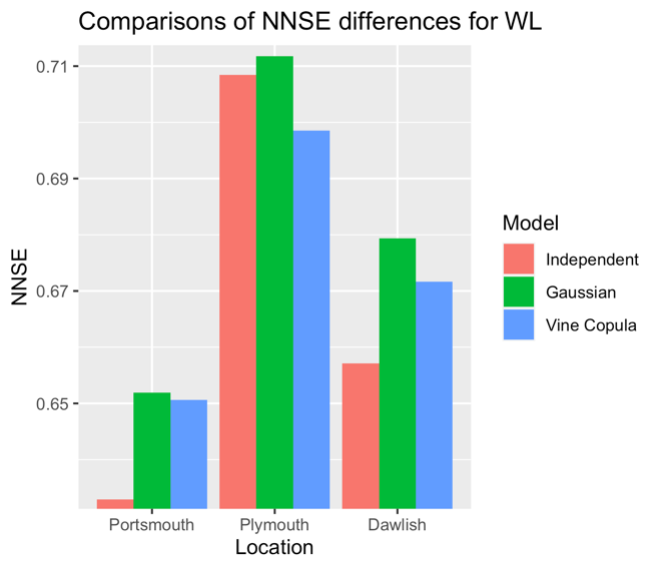} 
    \caption{Grouped bar charts showing the differences between optimal fit for each model and the NNSEs for \texttt{Hs} (left panel) and \texttt{WL} (right panel) for each location. The red bars show the results assuming independence among variables, the green bars assuming all Gaussian pair-copulas and the blue bars assuming a vine copula model. Shorter bars indicate better fitting models.}
    \label{fig:NNSE}
\end{figure}

Figure \ref{fig:NNSE} depicts grouped bar charts showing the differences between optimal fit for each model and the NNSEs for wave height (left panel) and water level (right panel) for each location. The red bars show the results assuming independence among variables, the green bars assuming all Gaussian pair-copulas and the blue bars assuming a vine copula model. Shorter bars indicate better fitting models. In the Plymouth location, the vine copula achieves better results than the other two models for both Hs and WL. The Gaussian model performs best for Hs in the Portsmouth location. The independent model is selected for the remaining cases, particularly in Dawlish, where again the lack of data points might be the cause. 

\begin{figure}[htbp]
    \centering
    \includegraphics[width=1\textwidth]{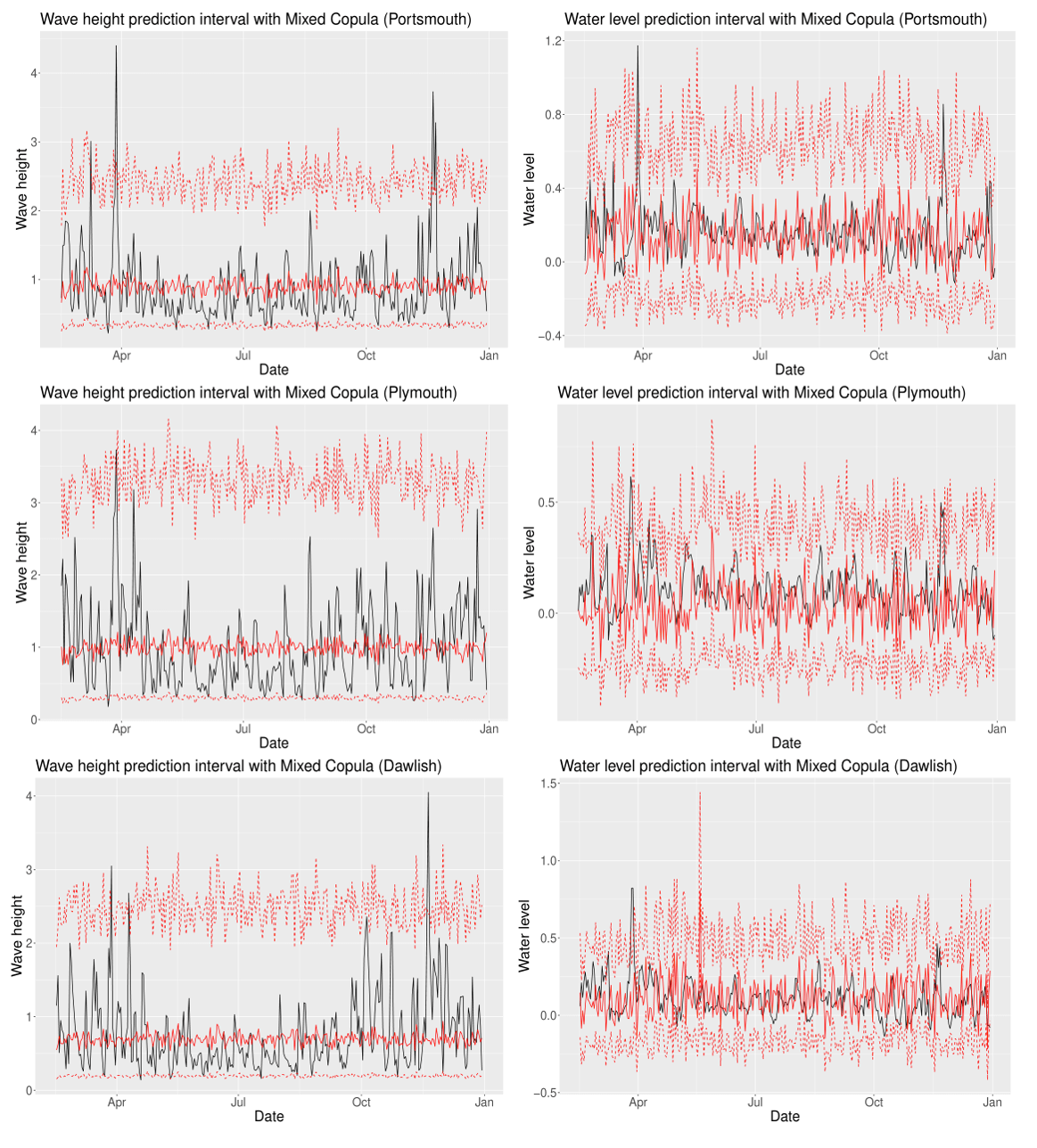} 
    \caption{Line plots showing forecasts and prediction intervals for \texttt{Hs} (left panels) and \texttt{WL} (right panel) obtained with the vine copula methodology for the period between the \nth{16} February 2016 and the \nth{31} December 2016, for Portsmouth (top panel), Plymouth (middle panel) and Dawlish (bottom panel). Observed values are in black, predicted values are the inner red lines and 95\% prediction intervals are the outer red lines.}
    \label{fig:oos_pred}
\end{figure}

Figure \ref{fig:oos_pred} shows the forecasts and prediction intervals for the wave height \texttt{Hs} and water level \texttt{WL} (on the left and right panel, respectively), obtained with the vine copula methodology for the period between the \nth{16} February 2016 and the \nth{31} December 2016.
The top panels depict the Portsmouth plots, the middle panels depict the Plymouth plots and the bottom panels depict the Dawlish plots.
The black lines denote the observed values, the inner red lines denote the predicted values and the outer red lines denote the 95\% prediction intervals.
We notice that the forecasted water levels are in line with the observations, and the average dynamics of wave height is adequately represented by the proposed model. Intervals predicted by the vine copula method capture most of the dynamic of the environmental variables, indicating that the proposed methodology is able to leverage social media information for forecasting flood-related data.  
\begin{figure}[htbp]
    \centering
    \includegraphics[width=0.45\textwidth]{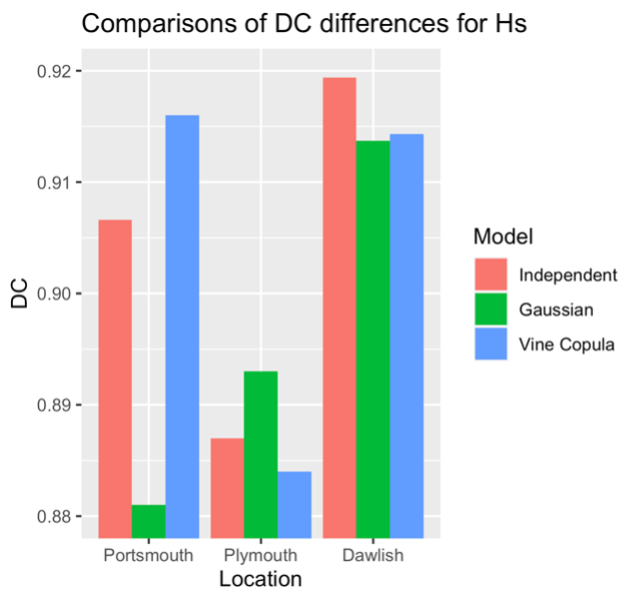} 
    \includegraphics[width=0.45\textwidth]{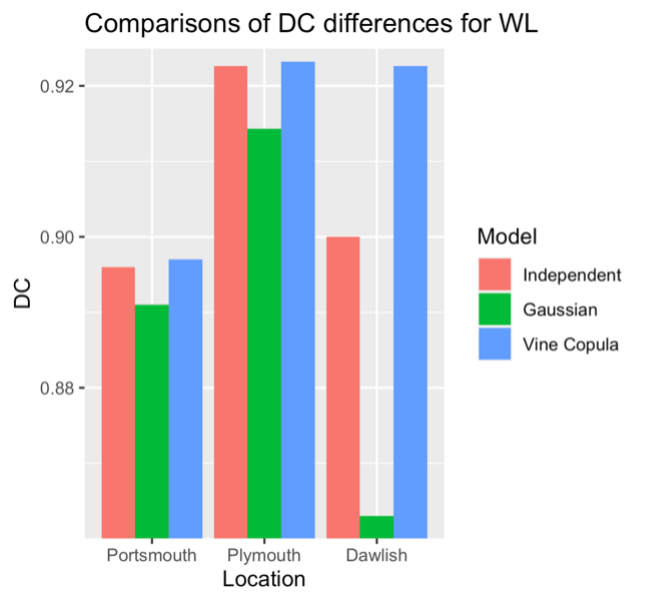} 
    \caption{Grouped bar charts showing the differences between optimal fit for each model and the DCs for Hs (left panel) and WL (right panel) for each location. The red bars show the results assuming independence among variables, the green bars assuming all Gaussian pair-copulas and the blue bars assuming a vine copula model. Shortest bars indicate better models.}
    \label{fig:DC}
\end{figure}

Figure \ref{fig:DC} illustrates grouped bar charts showing the differences between optimal fit for each model and the DCs for wave height (left panel) and water level (right panel) for each location. The bar colour codes are the same used in Figure \ref{fig:NNSE}. According to the DC, the Gaussian vine model is generally the preferred approach, while the vine copula model performs best for wave height in the Plymouth location. The independent model, which implies no input from the social media data for calculating predictions, is never selected.

%%%%%%%%%%%%%%%%%%%%%%%%%%%%%%%%%%%
%%      Conclusions		%%%
%%%%%%%%%%%%%%%%%%%%%%%%%%%%%%%%%%%

\section{Concluding Remarks}\label{sec:Conclusions}

In this paper, we propose a new methodology aimed at obtaining more accurate forecasts, compared to traditional approaches, for variables measuring inundations and floods events. The proposed methodology is based on the integration of environmental variables collected via remote sensing with online generated social media information. We obtained data at three different locations on the South coast of the UK, which were affected by severe storm events on several occasions in the past few years. Together with wave height and water level information, we also gathered Google Trends searches and Twitter microblogging messages involving keywords related to floods and storms. From the tweets, we considered the volume as well as the sentiment scores, to investigate the feelings of people towards inundation events. Our methodology is based on vine copulas, which are able to model the dependence structure between the marginals, and thus to take advantage of the association between social media and environmental variables. We tested our approach calculating out-of-sample predictions and comparing the vine copula method with two traditional approaches: the first based on a vine constructed with all Gaussian copulas, and the second based on independence between variables. The results show that the vine copula method outperforms the other two approaches in most cases, demonstrating that our methodology is able to leverage social media information to obtain more accurate predictions of floods and inundations than the other two approaches. In some cases, the Gaussian vine copula method is selected, showing that the vine data integration approach is still achieving the best performance, although some variables are less affected by asymmetries and tail dependence. Since social media information for Dawlish were lacking, they provided a more limited contribution to the prediction of the environmental variables for this location. 
The proposed methodology will support decision-makers enabling them to use knowledge gained from the model results to deepen their understanding of risks associated to floods and optimise resources in a more effective and efficient way. At strategic level, the methodology could be used to validate resource deployments in response to threats from floods; while at operational level, the methodology could assist to improve the effectiveness of civil contingency responses to flood events.

Further investigations involving other locations and including additional social media information will be the object of future work. Also, we will explore the use of the results of the study to validate inundation modes. Another extension will involve Bayesian inference, which would allow us to incorporate other information, such as experts’ opinion, in the model. In addition, the use of more sophisticated machine learning approaches could be envisaged for deriving the sentiment variables to improve the proposed methodology.

%%%%%%%%%%%%%%%%%%%%%%%%%%%%%%%%%%%
%%      Acknowledgements		%%%
%%%%%%%%%%%%%%%%%%%%%%%%%%%%%%%%%%%

\section*{Acknowledgements}

The authors are grateful to the anonymous Reviewers for their useful comments which significantly improved the quality of the paper. 
This work was supported by the European Regional Development Fund project \textit{Environmental Futures \& Big Data Impact Lab}, funded by the European Structural and Investment Funds, grant number 16R16P01302 .

%%     Bibliography 	%%%
%%%%%%%%%%%%%%%%%%%%%%%%%%%

\bibliographystyle{chicago}    
\bibliography{FSM}{}

\end{document}